\documentclass[11pt,a4paper]{article}
\usepackage{latexsym}
\usepackage{amsfonts}
\usepackage{amsmath}
\usepackage{epsf}

\author{Diego Meschini\footnote{E-mail: diego.meschini@phys.jyu.fi.} \qquad
Markku Lehto \qquad Johanna
Piilonen\footnote{E-mail: johanna.piilonen@phys.jyu.fi.} \smallskip\\
\emph{Department of Physics, University of Jyv\"{a}skyl\"{a}, } \smallskip \\ \emph{PL 35 (YFL), FI--40014 Jyv\"{a}skyl\"{a}, Finland.}}
\date{August 10, 2006}
\title{Geometry, pregeometry and beyond}

\begin{document}
\maketitle

\begin{abstract}
This article explores the overall geometric manner in which human
beings make sense of the world around them by means of their
physical theories; in particular, in what are nowadays called
pregeometric pictures of Nature. In these, the pseudo-Riemannian
manifold of general relativity is considered a flawed description
of spacetime and it is attempted to replace it by theoretical
constructs of a different character, ontologically prior to it.
However, despite its claims to the contrary, pregeometry is found
to surreptitiously and unavoidably fall prey to the very mode of
description it endeavours to evade, as evidenced in its
all-pervading geometric understanding of the world. The question
remains as to the deeper reasons for this human, geometric
predilection---present, as a matter of fact, in all of
physics---and as to whether it might need to be superseded in
order to achieve the goals that frontier theoretical physics sets
itself at the dawn of a new century: a sounder comprehension of
the physical meaning of empty spacetime.

\smallskip


\end{abstract}

 \tableofcontents

\section{Introduction} \label{Intro}
Quantum gravity is a field of physics whose specific subject
matter cannot be easily identified. In fact, one might even say
that there are almost as many conceptions of quantum gravity as
there are researchers of it. The task has been approached from innumerable angles, for different motivations, and with different underlying physical, mathematical, and philosophical assumptions. The tentative results are as varied. 

In very general terms, however, quantum gravity attempts to unify, in a physically meaningful way, the current understanding of Nature in terms of quantum mechanics and general relativity, which is as such perceived as incongruous and fragmented. Because, according to general relativity, gravitation is
associated with the geometric structure of spacetime, a common
factor of this search consists in reconsidering the structure of
spacetime from novel points of view. 

Among the numerous existing approaches, we have found a special
class rather conspicuous: those that attempt to go beyond the
pseudo-Riemannian manifold of physics without assuming any other
type of metric manifold for spacetime and without quantizing
general relativity taken as a presupposed starting point, but
rather accounting for spacetime in terms of entities, and possibly
their interactions, that are ontologically prior to it. This, we
believe, is a fair, \emph{preliminary} characterization of
\emph{what is known} by the name of \emph{pregeometry} (e.g.\
Monk, 1997, pp.\ 11--18).

Varied though pregeometric approaches can be, they all share a
feature in common: their all-pervading \emph{geometric
understanding} of the world. Such a kind of understanding is by no means particular to pregeometry; it is also present in all of
quantum gravity and, moreover, in all of physics. The intriguing,
particularly curious case that pregeometry makes with respect to
the previous assertion resides in its name and the corresponding
intentions of its practitioners, i.e.\ it strikes us as peculiar
to find geometric explanations in a field called, precisely,
pre-geometry. One must therefore pose the question: does
pregeometry live up to its name and matching objectives? And if
not, why not and in what sense not?\footnote{The criticism of
pregeometry to be put forward here is not based merely on a verbal
objection. It consists of an objection to the activities of
pregeometricians proper, given that they explicitly set out to
avoid geometry---and fail. Since this goal is, indeed, depicted
very well by this field's name, the criticism applies to the word
``pregeometry'' just as well. It is only in this sense that it is
\emph{also} a verbal objection.}

The clarification of the role of geometry in human physical
understanding and, in particular, in pregeometry---an attempt to
understand spacetime in a new light that tries to liberate itself
from geometry but to no avail---constitutes the content of the
present article. In Section \ref{Geometry}, the vital role of
geometry in all of physics and, in particular, in modern physics
will be discussed as preparation for the material ahead. In
Section \ref{Motivations}, the traditional, as well as the present
article's, motivations for the search for more novel structures
for spacetime will be reviewed. In Section \ref{PG}, pregeometry
and some of its schemes will be analyzed. Finally, in Section
\ref{BG}, a seldom investigated approach to the problem of
spacetime structure will be presented: the key to rethinking
spacetime anew might lie not in overtly geometric or pregeometric
explanations but in those that manage to go beyond geometry
altogether.

\section{Geometry and its role in physics} \label{Geometry}
The very conceptual foundations of geometry could be laid down in
the following way. Geometry and geometric thinking are built out
of two basic ingredients. The first and more primitive of them is
that of \emph{geometric objects} such as, for example, point,
line, arrow, polygon, sphere, cube, etc. As to the origin
of geometric objects, it appears after some consideration that
they must arise in no other way than through idealization by the
mind of sensory perceptions. Different natural objects are readily idealized by the human mind as possessing different shapes, and thus they become geometric objects of human thought. This activity of abstracting form from Nature, we believe, must have had its origin in the mists of antiquity.

The second and much more advanced ingredient that geometry
consists of is that of \emph{geometric magnitudes}. As its name
suggests, these magnitudes involve the expression of a quantity,
for which the concept of number is a prerequisite. Assuming that
the notion of counting is already available, geometric magnitudes
arise as result of a process (measurement) by means of which
geometric objects can be attributed different notions of
\emph{size}. For example, the geometric magnitude \emph{length}
applies only to a geometric object (e.g.\ line, curve) or a part
of it (e.g.\ edge of a polygon), and its original meaning is that
of counting how many times a unit of measure can be juxtaposed
along the object in question. Other geometric magnitudes are:
distance, area, volume, overlap, etc.

We stress that the said idealization of the natural object (and of
the physical unit of measure) as a geometric object is essential
here; without such a geometric abstraction, it would not be
possible to know in what way successively to lay down the standard
upon the measured object. In other words, the introduction of
geometric magnitudes presupposes the existence of geometric
objects.

Beyond the human primeval recognition of geometric form and, much
later, magnitude, the rise of more sophisticated geometric ideas
is recorded to have had its origin in the Neolithic Age in
Babylonia and Egypt. A landmark in the development of abstract
geometry, and perhaps the oldest geometric construction that rates
higher in abstraction than other common uses of geometric thought
is so-called Pythagoras' theorem. Incontestable evidence of
knowledge of Pythagorean triples\footnote{A Pythagorean triple
consists of an ordered triple $(x, y, z)$, where $x$, $y$, and $z$ satisfy the Pythagorean theorem $x^2+y^2=z^2$.} and of the Pythagorean theorem can be found in the Babylonian cuneiform text Plimpton 322 and in the Yale tablet YBC 7289, which date from between 1900 and 1600 b.C.\ in the Late Neolithic or Early Bronze Age.

There exist records, such as the papyrus Berlin 6619 dating from around 1850 b.C., showing knowledge of Pythagorean triples by
the Egyptians, too. However, no triangles are mentioned here or elsewhere. Van der Waerden (1983, p.~24) suggested that the Egyptians may have learnt about Pythagorean triples from the Babylonians. In support of this view, Boyer and Merzbach (1991) wrote: ``It often is said that the ancient Egyptians were familiar with the Pythagorean theorem, but there is no hint of this in the papyri that have come down to us'' (p.~17).

As far as the use of geometry is concerned, little has changed in
the last 4,000 years. Physical theories make extensive use of
qualitative and quantitative geometric concepts; moreover, in
trying to find a physical theory that does not include them, one
is certainly bound to fail: the role of geometry in all of past
and present physics is ubiquitous. At first glance, it appears
that especially quantitative geometric notions, such as sizes of
objects and distances between them, are only introduced into
theories to establish a connection with the world of experience
through measurements. However, just as significant is the fact
that great facilitation is achieved in human thought processes by
the introduction of geometric notions of both kinds. Probability,
a concept in principle free from any geometric connotations, will
be seen to be geometrized in this sense in Section \ref{GQMec}.

An example from another scientific field will be given presently
to clarify the second reason above. In the theory of coding, one
is concerned with sending messages over a noisy information
channel and with being able to recover the original input from a
possibly distorted output. Evidently, no intrinsically geometric
magnitudes are involved in this problem; however, these are
readily constructed to ease analysis and understanding of the
situation. A word $c$ is thought of as a vector $\vec
c=(c_0,c_1,\ldots,c_{n-1})$ and a distance function, called
Hamming metric, is defined for any pair of words $c$ and $c'$ as
$\max\{i\ |\ c_i\neq c'_i, 0\leq i\leq n\}$. After certain
assumptions about the nature of the channel, a received word is
corrected searching for the codeword that is \emph{closest} to it
in the above sense, i.e.\ differing from a codeword in the least
number of elements. As this example shows---and as will become
progressively clearer---it would seem it is a trait of the human
mind to seek a geometric understanding of the world.

In physical contexts,\footnote{This is not necessarily so in
mathematical contexts where, for example, norm and distance can
arise from non-inner-product constructions such as
$\|f\|=\sup_{x\in \mathbb{R}}|f(x)|$ in Banach spaces, and
$d(x,y)=\min_{\mathrm{paths}(x,y)}\{ n(\mathrm{edge}) \}$ in graph theory.} there exists the more primitive concept of inner product from which three key, quantitative geometric notions emerge, namely: overlap of $|A\rangle$ and $|B\rangle$, $\langle
A|B\rangle$ (with parallelism and orthogonality arising as special
cases of this concept); length of $|A\rangle$, $\langle
A|A\rangle^{1/2}$; and distance\footnote{If the space to which
$|A\rangle$ and $|B\rangle$ belong were not flat, one should then
resort to a line integral to find out this distance; yet, the
essence of the concept remains the same (see Section
\ref{GGRel}).} between $|A\rangle$ and $|B\rangle$, \mbox{$\langle
A-B|A-B\rangle^{1/2}$}. Therefore, the \emph{inner product} has the
power of introducing the most significant quantitative geometric
notions into physical theories.

In what follows, the geometric way Nature is understood in modern
times by means of the theories of general relativity and quantum
mechanics will be investigated. It will be seen that, although the
portrayal of the physical world has come a long way since the dawn
of mankind, something has remained unchanged in it ever since: the
everywhere-present use of geometric explanations. Although dressed
up in much more sophisticated garments, the role of geometry in
today's received description of the world is, just as in earlier
times, nothing short of crucial. This is contrary to common
belief, as expressed in Ashtekar's statement:
\begin{quote}
[W]e can happily maintain a \emph{schizophrenic} attitude and use
the precise, \emph{geometric picture of reality} offered by
general relativity while dealing with cosmological and
astrophysical phenomena, and the quantum-mechanical world of
chance and intrinsic uncertainties while dealing with atomic and
subatomic particles. (Ashtekar, 2005, p.~2) [Italics added]
\end{quote}

\subsection{Geometry in general relativity} \label{GGRel}
According to general relativity, spacetime is described by means
of a curved, four-dimensional manifold $M$. But what objects
characterize spacetime physically? At first glance, these would
seem to be spacetime's building blocks, its points. On closer
inspection, it can be noticed that it is not the points that are
physically real but the measurable interval relations
$\mathrm{d}s^2$ between point-events $E$, $F$, etc. Specifying now
these intervals, i.e.\ the generalized distances between
neighbouring point-events, the geometry of spacetime is fixed.
This is in accordance with the earlier remarks that quantitative
geometry arises from a specification of sizes of objects or
distances between them.

Misner, Thorne, and Wheeler (1973, pp.~306--308) analyzed the
immensity of the problem of giving the distances between a vast
numerousness of point-events by crudely writing down a number for
each pair. They showed how the method can be refined until all
that needs to be provided is a function
$\mathbf{g}(\overline{E-F},\overline{E-F})$ capable of producing
the squared interval between neighbouring events $E$ and $F$. On
the basis of the metric, now the distance between any two
point-events in spacetime arises as the line integral, along the
shortest joining curve $\sigma$, of the (infinitesimal) length of
the tangent vector $\vec T_\sigma$ to $\sigma$ at each point;
$\int_\sigma \mathbf{g}^{1/2}[\vec T_\sigma  (u),\vec T_\sigma
(u)]\mathrm{d}u$. Hence the introduction of a metric field
$\mathbf{g}(x)$ as a local concept in general relativity.

Mathematically, the metric is a two times covariant tensor field
$\mathbf{g}(x)$ equivalent in every respect to an \emph{inner
product} $\langle\ |\ \rangle$ of vectors on the tangent space
$T_P(M)$ of the spacetime manifold $M$ at every point $P$. Even if
one knew nothing of general relativity and were at the moment
totally ignorant of the purposes of the existence of an inner
product in this theory, it could be safely taken, as emphasized
earlier, as a sign of the introduction of quantitative geometry
into the theory. This is because the inner product is an
undisputable bearer of geometric magnitudes. In the theory, moreover, geometric objects (point, geodesic, tangent plane, arrow vectors, etc.) are everywhere present.

In dealing with spacetime as a mathematical manifold $M$ as we
just did, the association is made between physical point-events
$E$ of spacetime and mathematical points $P$ of $M$. This
association is extremely significant and potentially perilous, and
must not be taken lightly; whereas point-events are endowed with a
perfectly clear physical meaning (say, a flash of lightning), the
case is not so for the spacetime points $P$ themselves. The
physical reality or unreality of the latter continues to be the
object of heated philosophical debate and constitutes, as it were,
one of the issues underlying our research.

All the geometric aspects of spacetime are \emph{logically} (but
not physically) obtained in terms of the metric tensor
$\mathbf{g}(x)$, the most important of these being, by far, the
invariant interval $\mathrm{d}s^2$, given by means of the
famous relation $\mathrm{d}s^2=g_{\mu\nu}(x)\mathrm{d}x^\mu
\mathrm{d}x^\nu$. The significance of this interval cannot be
overemphasized: it constitutes the most basic \emph{physical}
relationship between spacetime point-events $E$ that remains
unchanged for any system of reference in any state of motion,
while at the same time giving a physical basis for the metric
field.

Finally, the metric field also logically defines manifold
curvature, and thus gravitation acquires a \emph{new} geometric
character:\footnote{Notice that gravitation's Newtonian
description as a vectorial force was also a geometric concept,
namely, a directed arrow in Euclidian space with a further notion
of size (strength) attached.} free particles are now thought to
fall along spacetime geodesics. Therefore, it can be said
with little hesitation that the use of geometric concepts is
essential in general relativity.

\subsection{Geometry in quantum mechanics} \label{GQMec}
Imagine a plate which only transmits light polarized perpendicular
to its optical axis and a beam of light, polarized at an angle
$\theta$ with the above axis, to be fired at the plate such that
photons arrive one at the time. How is one to understand the
result that the intensity of the transmitted light is a fraction
$\sin^2(\theta)$ of the total intensity in
terms of \emph{individually polarized photons}? 
This experimental result can be understood thinking that the
polarization state $|P\rangle$ \emph{of each photon} can be
expressed as a linear superposition of two polarization states
$|P_t\rangle$ and $|P_a\rangle$, one associated with transmission
and the other with absorption, respectively: \mbox{$|P\rangle=\langle
P_t|P\rangle |P_t\rangle+\langle P_a|P\rangle |P_a\rangle$}, where
$\langle P_i|P\rangle$ ($i=t,a$) denotes ``how much'' states
$|P\rangle$ and $|P_i\rangle$ have in common. With the further
requirement that $\sum_i |\langle P_i|P \rangle|^2=1$, one can
associate $|\langle P_t|P \rangle|^2$ with the probability of an
individual photon being transmitted and $|\langle P_a|P
\rangle|^2$ with its being absorbed. When the number of photons is
large enough, the classical result is recovered (Dirac, 1958,
pp.~4--7).\footnote{For an illuminating account of the
experimental and now conventional theoretical foundations of
quantum mechanics, see Chapters 1 and 2 of this reference.}

This physically inspired superposition principle leads directly to
the suggestion that quantum-mechanical states consist of
\emph{metric vectors} or \emph{arrows},\footnote{The inner product
effectively turns in principle shapeless state vectors into
arrows, since geometric magnitudes demand geometric objects.}
since it holds physically that the linear combination of two
states is also a state, and the overlap (cf.\ \emph{inner
product}) of two states gives the probability amplitude of the
original state collapsing onto another. Despite being thus
inspired in the natural world, however, the situations that state vectors
describe evade any mechanistic interpretation (i.e.\
visualization), and the physical meaning of state vectors
themselves continues to incite a great deal of philosophical
controversy.

State vectors belong to a Hilbert space $\mathcal{H}$, whose
inner-product structure is vital in order to sustain quantum
theory. This is because the notion of probability wants an inner
product after the geometrically minded manner in which quantum
mechanics naturally developed. The formalism of quantum mechanics postulates that, concerning an
observable $O$ and a system in state $|\psi\rangle$, the
probability that the degenerate eigenvalue $\lambda_i$ with
associated eigenvectors $|u_{i,j}\rangle$ will result is
$P(\lambda_i)=\sum_{j=1}^{g_i} |\langle u_{i,j}|\psi\rangle|^2$.
In words, this probability is given by the sum of the squared
\emph{lengths} of the \emph{projections} of the original state
\emph{vector} onto the basis vectors of the target space. Thus
probability, which is in principle a concept that only has to do
with counting positive and negative outcomes, acquires its
geometric meaning.

In this connection, Isham (1995, pp.~13--17) explained how,
contrary to the case in statistical physics, in quantum mechanics
probabilities originate in Pythagoras' theorem in $n$ dimensions.
Given an initial state vector $|\psi\rangle$, the interest is in
predicting what are the chances that it will turn into a certain
other after measurement. Then, if all the possibly resulting state
vectors are organized at right angles with each other and each at
an angle $\theta_i$ with $|\psi\rangle$, it will hold that
$\sum_{i=1}^n \cos^2(\theta_i)=1$ by virtue of Pythagoras' theorem.
Finally, the association of each term with a probability of a
particular result immediately suggests itself.

It is in this manner that probability is geometrized in order to
provide the human mind with geometric understanding of, in
principle, geometry-unrelated observations. Pythagoras' theorem
comes thus from its humbler, Neolithic uses to serve human
understanding in yet another facet of Man's inquiries into Nature.

\section{Rethinking spacetime} \label{Motivations}
It is currently believed that the spacetime manifold of general
relativity is in need of rethinking, pregeometry being one of the
most drastic means to this end. Two paradigmatic motivations
behind this search are the various forms of gravitational collapse
predicted by general relativity and the confirmed existence of
quantum-mechanical correlations.\footnote{See Butterfield and
Isham (2000, pp.\ 34--36) for a classification of motivations according to whether they are viewed from the perspective of the particle physicist or the general relativist.} In fact, Monk (1997, pp.~3--9) presented these two points as central for the search for
a quantum theory of gravity; yet we believe that their relevance
to this search is not necessarily guaranteed. We present our own
perspective in Section \ref{PEST}.

\subsection{Spacetime manifold collapse}
Misner et al.\ (1973) called gravitational collapse ``the greatest
crisis in physics of all time'' (p.~1196). The singularities
predicted by general relativity, it is argued, are events at which
the pseudo-Riemannian manifold that characterizes spacetime breaks
down. This is seen as a sign that spacetime only manages to look
like a continuous manifold but that, in fact, it may be better
represented by a theoretical picture of a different kind. But do
these spacetime singularities represent a problem, or are they
simply irrelevant cases where general relativity has reached its limits of
applicability, much like the Newtonian gravitational law
$-(GMm/r^3)\vec r=m(\mathrm{d}^2\vec r/\mathrm{d}t^2)$ when $r=0$?
In this light, general relativity could be taken as a complete and
consistent theory of spacetime, singularities being an inconsequential domain outside it.

At this point, one must not necessarily discard spacetime
singularities as motivations for the search for better theories.
However, such singularities may only be taken as proper
motivations if one believes that they are in fact meaningful
questions whose solution will have a bearing on the understanding
of the newer theories. This might or might not be the case. The
case of Newtonian singularities, for their part, was certainly not
relevant to the invention of relativistic or quantum theories. On
the other hand, the problem of electromagnetic singularities was
relevant to the development of a quantum-mechanical theory, as
evidenced by Bohr's efforts to solve the problem of the collapse
of the atom.

\subsection{Failure of locality}\label{QMC}
Stapp (1975) called Bell's (1987) theorem ``the most profound
discovery of science'' (p.~271). Indeed, another paradigmatic
motivation to rethink the description of spacetime stems from the
analysis of non-local, quantum-mechanical correlations, whose
experimental confirmation (e.g.\ Aspect, Dalibard, \& Roger, 1982) tells
that two distant parts of an entangled system know more about
each other than local, realistic premises
allow.\footnote{Locality forbids signals to propagate
instantaneously in space, and realism is meant in the sense that
systems possess values for their physical magnitudes independently
of their being observed.} Whereas some have, in response, searched
for an explanation within the spacetime manifold
framework,\footnote{See e.g.\ (Selleri, 1988, pp.~vii--ix) for a list of alternatives.} others
have maintained that, since non-locality is not well-integrated in
the manifold notion of spacetime, space is not what it is believed
to be; that, although the correlated subsystems are irreconcilably
distant within the traditional manifold notion of space, they may
as well present a more amenable relationship---in this
respect---within a different framework in which space is not a
manifold, but a structure of a different kind.\footnote{See
Section \ref{EJ} and references therein for a proposed solution of
this type. In particular, Eakins and Jaroszkiewicz (2003) wrote:
``If Einstein locality is synonymous with classical Lorentzian
manifold structure, and if this structure is emergent, then it
seems reasonable to interpret quantum correlations as a signal
that there is a pre-geometric (or pre-emergent) structure
underlying the conventional spacetime paradigm'' (p.~17).}

However, since quantum non-locality may as well be taken as an
unproblematic, specific feature of quantum mechanics, a similar
criticism as above ensues. Again one cannot tell whether these
correlations that defy the ideas of locality should be addressed
as a troublesome feature to be accommodated, a problem to be
solved, by a theory that will supersede the current ones, or
whether they simply are an intrinsic, trouble-free characteristic
of quantum mechanics irrelevant to any future developments. As
with spacetime singularities, this criticism does not aim at
denying the significance of quantum non-locality as a motivation
for research on the structure of space and time, but only to put
it in its proper perspective.

\subsection{The physical existence of space and time} \label{PEST}
It is here put forward that a more substantial reason for
investigating these matters is to try to find out whether space
and time might have an own structure independently of matter and
fields; in other words, whether \emph{empty} spacetime can claim
physical existence. Despite the phrase being a commonplace, the meaning of ``empty space'' or ``empty spacetime'' is far from trivial. Does the phrase refer to spacetime as devoid of all matter, or also devoid of \emph{all} fields? 

Due to its particular feature of affecting
all free bodies in the same way, gravitation has become a
geometric property of spacetime, its curvature, and is logically
characterized by spacetime's metric field $\mathbf{g}(x)$.
Therefore, from this perspective, gravitation is always present as
a \emph{content} of any metric spacetime. One is inescapably given
with the other, even if the spacetime in question were a flat,
Minkowskian one. This conception is, in fact, Einstein's:
\begin{quote}
If we imagine the gravitational field, \emph{i.e.}\ the functions
$g_{ik}$, to be removed, there does not remain a space of the type
(1) [flat], but absolutely nothing\ldots A space of the type (1),
judged from the standpoint of the general theory of relativity, is
not a space without a field, but a special case of the $g_{ik}$
field\ldots There is no such thing as an empty space, \emph{i.e.}\ a space without field. Space-time does not claim existence on its own, but only as a structural quality of the field. (Einstein, 1952b, p.~176)
\end{quote}
As a consequence, in order to vacate spacetime, one must, along
with all its matter, rid spacetime of its \emph{quantitative
geometry}. When the metric field is gone, so is the network of
intervals $\mathrm{d}s^2$ logically built upon
it,\footnote{Intervals $\mathrm{d}s^2$ here refer to relations
between \emph{points}. Intervals between physical events would not
go anywhere if the metric were to disappear so long as we still
had a geometrodynamic clock (Ohanian, 1976, pp.~192--198)---\emph{physically}, intervals between
events are prior to the metric field.} and only spacetime's
building blocks remain: its points. However, according to general
relativity's diffeomorphism invariance principle, these points
cannot be observed by means of any physical tests, and empty
spacetime---so the argument remains---cannot claim physical
existence.

Must one give up on the problem of the existence of empty
spacetime at the point where a deadlock seems to have been
reached? Not quite, because in the above quantum theory---so far
the only branch of natural science forced to confront directly the
problem of physical \emph{existence}\footnote{See (Isham, 1995,
p.~65).}---was not taken into account. Moreover, we propose that
understanding spacetime anew demands going beyond its intrinsic
geometric description \emph{entirely}; not only the metric (source
of geometric magnitudes) but also the points (geometric objects)
must be left behind. For this task, the use of \emph{non-geometric means} should be required, as well as the guidance of some \emph{physical principle} directly relevant to the \emph{existence of spacetime}, so that one's search would not be blind guesswork.

One the one hand, the choice to transcend geometric methods is
based on the very general observation that genuine scientific
explanations come about only when older concepts are explained in
terms of new ones of a different kind. Spacetime, from its metric
structure down to its points, is first and foremost a geometric
concept; as such, it may require a more basic explication of a
non-geometric kind. History teaches us that progress in the
scientific world-view is achieved only when such fundamentally
different explanations are given for the theoretical notions in
current use (e.g.\ the new understanding of gravity as the
curvature of spacetime). Might this be a hint that a better
understanding about the nature of spacetime could be achieved by
going beyond its geometric bases altogether? One the other hand,
the need to supersede geometry entirely also transpires, in our
view, from a look at the stalemates that physicists and
philosophers, in their need for familiar geometric concepts and
words, continue to relentlessly confront in their study of the
ontology of space, time, and the quantum principle; most notable
among these problems is the search for the physical meaning of
spacetime points, state vectors, and related issues.

The choice invariably made by practically everyone---and as shall
be seen next, unwittingly, by pregeometricians---is to attempt to
clarify the geometric nature and structure of the spacetime
manifold with more geometric concepts. The choice to explore
spacetime from a non-geometric point of view favoured here is a
drastic alternative and a path that hardly anyone (but for
Clifford, Eddington and Wheeler; see Section \ref{BG}) has
attempted to walk before; and certainly one that no-one has walked
to completion.

\section{Pregeometry} \label{PG}
The word ``pregeometry'' is evocative of positioning oneself
before geometry. But what does this mean in more detail? In the
Introduction, a characterization of pregeometry
was given as an attempt, by design, to account for
spacetime geometry resting on entities ontologically prior to it
and of an essentially new character. However, pregeometry has
turned out to be an incongruous enterprise. Indeed, the
distinctive feature of pregeometric approaches is that they drop
the assumption of spacetime being correctly described by a
pseudo-Riemannian manifold; but \emph{only to replace it with some
type or other of geometric-theoretical conception}.

Nothing supports this statement more dramatically and vividly than
the following quotation from Gibbs:
\begin{quote}
Once it has been decided which properties of space-time are to be
discarded and which are to remain fundamental, the next step is to
choose the appropriate \emph{geometric structures from which the
pregeometry is to be built}. (Gibbs, 1996, p.~18) [Italics added]
\end{quote}
In this use of the term, pregeometry is synonymous to
pre-Riemannian-manifold physics, although not synonymous to
pregeometry in the proper semantic and historic sense of the
word---the sense its inventor, J.~A.~Wheeler, endowed it with and
which its current practitioners attempt to fulfill; in terms of
Wheeler's (1980) two most striking pronouncements: ``a concept of
pregeometry that breaks loose at the start from all mention of
geometry and distance'' and ``\ldots to admit distance at all is
to give up on the search for pregeometry'' (p.~4) (See Section
\ref{WP}).

In connection with these quotations from Wheeler and with those
from Gibbs above, we also find the following remarks by Cahill and
Klinger quite striking and worthy of mention:
\begin{quote}
The need to construct a non-geometric theory to explain the time
and space phenomena has been strongly argued by Wheeler, under the
name of pregeometry. Gibbs has recently compiled a literature
survey of such attempts. (Cahill \& Klinger, 1996, p.~314)
\end{quote}
At the reading of this passage, one is at first quite pleased and
subsequently nothing but disheartened. For, whereas it is true
that Wheeler vowed for a non-geometric theory of space and time
and named it pregeometry, it could not be farther from the truth
to say that the pregeometric theories devised so far (for example,
those in Gibbs' or Monk's survey) refrain from the use of
geometric concepts; to the contrary, they use them extensively, as
is our purpose to show.

In order to support the above statements, an analysis of some
pregeometric works will be presented shortly, in which
pregeometry's modes of working will be investigated from the
particular perspective of this article. What will follow is
therefore not an impartial review, nor does it intend to be
complete in its coverage of its subject-matter but only to display
some show-pieces of it. At the risk of displeasing some authors,
attempts to transcend the spacetime manifold in the way explained
earlier are here classified as pregeometric, even when this
designation is not used explicitly in their programmes. This way
of proceeding is somewhat unfair since the following works will be
criticized mainly as far as they are to be expressions of a
\emph{genuine pregeometry}. When they exist, explicit mentions of
pregeometry will be highlighted; lack of any such mentions will be
acknowledged as well.  Broad categories of pregeometric
frameworks, although possibly mutually overlapping, will be
identified. One or more actual cases exemplifying each group will
be included.

\subsection{Discreteness} \label{Disc}
One category could have embraced most of---if not all---the works
in this analysis. Without doubt, the best-favoured type of
programme to deal with new ways of looking at spacetime is, in
general terms, of a discrete nature. The exact meaning of
``discrete'' we will not attempt to specify; the simple intuitive
connotation of the word as something consisting of separate,
individually distinct entities (Merriam-Webster Online Dictionary)
will do. A choice of this sort appears to be appealing to
researchers; firstly, because having cast doubt upon the
assumption of spacetime as continuous, its reverse, discreteness
of some form, is normally envisaged; secondly, because
quantum-theoretical considerations also tend to suggest, at least
on an intuitive level, that also spacetime must be built by
discrete standards.

Last but not least, many find some of the special properties of
discrete structures quite promising themselves. Among other
things, such structures seem to be quite well-suited to reproduce
the continuous manifold in some limit, because they
\emph{naturally support geometric relations} as a built-in property. As
is well-known, this was originally Riemann's realization. He went
further in his remarks as to the possible need for such a type of
structure:
\begin{quote}
Now it seems that the empirical notions on which the metrical
determinations of space are founded, the notion of a solid body
and of a ray of light, cease to be valid for the infinitely small.
We are therefore quite at liberty to suppose that the metric
relations of space in the infinitely small do not conform to the
hypothesis of geometry; and we ought in fact to suppose it, if we
can thereby obtain a simpler explanation of phenomena.

The question of the validity of the hypotheses of geometry in the
infinitely small is bound up with the question of the ground of
the metric relations of space. In this last question, which we may
still regard as belonging to the doctrine of space, is found the
application of the remark made above; that in a discrete
manifoldness, the ground of its metric relations is given in the
notion of it, while in a continuous manifoldness, this ground must
come from outside. Either therefore the reality which underlies
space must form a discrete manifoldness, or we must seek the
ground of its metric relations outside it, in the binding forces
which act upon it. (Riemann, 1873, p.~17)
\end{quote}

Riemann can therefore be rightly considered the father of all
discrete approaches to the study of spacetime structure. He should
not be considered, however, the mentor of an authentic pregeometry
since, as he himself stated, discrete manifolds naturally support
geometric relations, becoming, through these relations, geometric objects themselves.\footnote{Geometric objects can appear either self-evidently, or first as general, abstract objects, which subsequently become geometric objects via the attachment of geometric magnitudes to them. For example, a ``set of objects with neighbouring relations'' becomes a graph, i.e.\ a set of \emph{points} with joining \emph{lines}, when a distance
is prescribed on the abstract relations. In quantum mechanics, we saw that state vectors likewise become geometric objects (arrows) after the specification of their inner product.}
The absolute veracity of Riemann's statement can be witnessed in every attempt at pregeometry, not as originally designed by Wheeler, but as it of itself came to be.

\subsection{Graph-theoretical pictures} \label{GP}
Graphs are a much-favoured choice in attempts at pregeometric
schemes. Following the characterization made by Wilson (1985, pp.\
8--10 and 78), a graph of a quite (although perhaps not the most)
general type consists of a pair $(V(G),E(G))$, where $V(G)$ is a
possibly infinite set of elements called vertices and $E(G)$ is a
possibly infinite family of ordered pairs, called edges, of not
necessarily distinct elements of $V(G)$. Infinite sets and
families of elements allow for an infinite graph which can be,
however, locally-finite if each vertex has a finite number of
edges incident on it. Families (instead of sets) of edges permit
the appearance of the same pairs more than once, thus allowing
multiple edges between the same pair of vertices; in particular,
an edge consisting of a link of a single vertex to itself
represents a loop. Finally, ordered (rather than unordered) pairs
allow for the existence of directed edges in the graph.

A particular reason for the popularity of graphs in pregeometry is
that, along the lines of Riemann's thoughts, they naturally
support a metric. It is especially noticed here that, when a
metric is introduced, all edges are taken to be of the same length
(usually a unit) irrespective of their actual shapes and lengths
as drawn on paper or as conjured up by the imagination. This
distinguishes graphs from certain types of lattices, which have a
more literal interpretation, as will be seen in Section \ref{LP}.

\subsubsection{Dadi\'c and Pisk's discrete-space structure}
Dadi\'{c} and Pisk (1979) went beyond the manifold assumption by
representing ``space as a set of abstract objects with certain
relations of neighbourhood among them'' (p.~346), i.e.\ a graph;
in it, the vertices appear to correspond to space's points. This
approach assumes a \emph{metric} as naturally inherited from the
graph, and a definition of dimension---based on a modification of
the Hausdorff dimension---that is scale-dependent. The graph
$|G\rangle$ representing spacetime is required to be an unlabelled
one in its points and lines and must be characterizable by just
its topological structure. Operators $b^\dagger$ and $b$ are
defined for the creation and annihilation of lines, such that
$|G\rangle$ can be constructed from the vacuum-state graph
$|0\rangle$ by repeated application of $b^\dagger$; this is indeed
a Fock-space framework.

What led Dadi\'c and Pisk to this choice of graph-theoretical
approach? In this respect, they argued that they found the
existence of some objects and the relation of neighbourhood
between them to be essential in the intuitive notion of space.
Based on this, they proceeded to construct their discrete-space
structure. However, it can be realized that this basis is not
grounded on any further physical principles.

Quantum mechanics---in particular, a Fock-space method---is used
in this approach, if nothing else, as a formalism to deal with
graphs, their states, and their evolution. One notices at this
point that Fock spaces include a metric as part of their built-in
structure, too.

As can be seen, geometric concepts are here clearly assumed; they
were typified by the natural metric defined on the graph on the
one hand, and that of Fock spaces on the other, and not least by a
graph as a composite geometric object (vertices and edges) itself.
This fact alone would make this approach a questionable case of an
authentic pregeometry, although it must be acknowledged that its
authors did not so argue.

\subsubsection{Antonsen's random graphs}
Antonsen (1992) devised a graph-theoretical approach that is
statistical in nature. Space is identified with a dynamical,
random graph, whose vertices appear to be associated with space
points and whose links are taken to have unit \emph{length}; time
is given by the parameterization of these graphs (spaces) in a
metaspace. Primary point creation and annihilation operators
$a^\dagger$ and $a$ are defined, as well as secondary
corresponding notions for links, $b^\dagger$ and $b$;
probabilities for the occurrence of any such case are either given
or in principle calculable.

This approach is, according to its author, ``more directly in tune
with Wheeler's original ideas'' (p.~8) and, in particular,
similar to that of Wheeler's ``law without law'' (p.~9) in that
the laws of Nature are viewed as being a statistical consequence
of the truly random behaviour of a vast number of entities working
at a deeper level. One must have reservations about this view
since Antonsen is already working in a geometric context (a graph
with a notion of distance), which is not what Wheeler had in mind.
Moreover, there does not appear to be any principle leading
Antonsen to introduce this kind of graph structure in the first
place. He sometimes (pp.~9, 84) mentioned the need to assume as
little as possible, perhaps overlooking the fact that not only the
quantity but also the nature of the assumptions made---as few as
they might be---bears as much importance.

According to Antonsen, this framework does not really assume
traditional quantum mechanics although it might look like it. He
claimed that the operators and the Hilbert space introduced by him
are only formal notions without direct physical interpretations
besides that of being able to generate a geometric structure. On
the contrary, he attempted to derive quantum mechanics as a
hidden-variable theory.

Antonsen is therefore the first claimant to pregeometry who calls
for the use of geometric objects (a graph) and magnitudes
(distance) in its construction; for this reason, his scheme is
rather suspicious as an authentic expression of pregeometry.

\subsubsection{Requardt's cellular networks}
Requardt also went beyond the manifold assumption by means of a
graph-theoret\-ical approach. Space is associated, at a deeper
level, with a graph whose nodes $n_i$ are to be found in a certain
state $s_i$ (only differences of ``charge'' $s_i-s_k$ are
meaningful), and its bonds $b_j$ in a state $J_{ik}$ that can be
equal to 0 or $\pm 1$ (vanishing or directed bonds, respectively).
A law for the evolution in what he called the ``clock-time'' of
such a graph is simply introduced (Requardt \& Roy, 2001,
pp.~3042--3043) on the grounds that it provides, by trial and
error, some desired consequences. Requardt (2000, p.~1) then
assumed that this graph evolved from a chaotic initial phase or
``big bang'' in the distant past, characterized by the complete
absence of stable patterns, in such a way as to have reached a
generally stable phase---explained by the theory---which can be
associated with ordinary spacetime.

A peculiarity of this scheme is that it seeks to understand the
current structure of spacetime as emergent in time,\footnote{As is typical of arguments considering the dynamic emergence of time (also as in space\emph{time}), it is unclear in what dynamic time the emergent time is supposed to appear. This hypothetical, deeper dynamic time, when considered at all (discrete ticks of clock time for Requardt), is nevertheless always the same kind of simple and intuitive external parameter as the notion of time it attempts to give rise to; the end result is to trade one little-understood, external-parameter time for another.} i.e.\ as a consequence of the
primordial events that are believed to have given rise to it in
the first place. In this respect, it could have also been
classified as a cosmological scheme, although not as a
quantum-cosmological one since quantum
mechanics is not assumed. Regarding this issue, Requardt (1996)
explained that his goal is to ``identify both gravity and quantum
theory as the two dominant but derived aspects of an underlying
discrete and more primordial theory\ldots'' (pp.~1--2).

It is interesting to note that, in contrast to the preceding
scheme, in this scheme spacetime points are not associated with
the graph's vertices but with ``densely entangled subclusters of
\emph{nodes} and \emph{bonds} of the underlying network or
graph.'' More specifically, they are cliques, ``the maximal
complete subgraphs or maximal subsimplices of a given graph''
(Requardt \& Roy, 2001, p.~3040).

A lack of guiding principles must again be criticized. Requardt's
(1995) choice of approach seems to rest on ``a certain suspicion
in the scientific community that nature may be `discrete' on the
Planck scale'' (p.~2). Therefore, he argued, cellular networks
make a good choice because they are naturally discrete and can,
moreover, generate complex behaviour and self-organize. However,
it must be said that basing a choice on a generally held suspicion
whose precise meaning is not clear may not be very cautious.

Although Requardt (1995) intended to produce a pregeometric scheme
(``What we are prepared to admit is some kind of
\mbox{`\emph{pregeometry}'\ ''} (p.~6)) and strove for ``a
substratum that does not support from the outset\ldots geometrical
structures'' (p.~2), he soon showed---in total opposition to the
earlier remarks---that his graphs ``have a natural metric
structure'' (p.~15). He introduced thus a very explicit notion of
\emph{distance}. Requardt was concerned with avoiding the
assumption of a continuous manifold, and in this he succeeded.
However, geometric notions are not merely manifold notions, as
argued earlier. This attempt at pregeometry suffers therefore from
the same geometric affliction as the previous one.

\subsection{Lattice pictures} \label{LP}
At an intuitive level, (i) a lattice consists of a regular
geometric arrangement of points or objects over an area or in
space (Merriam-Webster Online Dictionary). From a mathematical
point of view, (ii) a lattice consists of a partially ordered set
$\langle L,\prec\rangle$ in which there exists an infimum and a
supremum for every pair of its elements.\footnote{A lattice can
also be defined as an algebra $\langle L,\wedge,\vee\rangle$, with
$\wedge$ and $\vee$ subject to certain rules, rather than as a
partially ordered set. It can be shown (e.g.\ Eric Weisstein's
World of Mathematics, http://mathworld.wolfram.com/Lattice.html)
that, after the identifications $a\wedge b=\inf\{a,b\}$, $a\vee
b=\sup\{a,b\}$ and $a\leq b \Leftrightarrow a\wedge b=a$, these
two concepts are equivalent.}

The first notion of lattice (i) goes hand in hand with the same
idea as used in Regge calculus. In it, an irregular lattice is
used in the sense of an irregular mesh whose edges can have
\emph{different} lengths. For this reason, such a lattice cannot
be a graph, where all links are equivalent to one another. The
second notion of lattice (ii) is, \emph{to a degree}, connected
with the approach to spacetime structure called causal sets.
Although strictly speaking, a causal set is only a locally finite,
partially ordered set without loops, when it also holds that there
exists a unique infimum and supremum for every pair of its
elements (existence of a common past and a common future), the
causal set becomes a lattice in the second sense of the term.
Despite the fact that they are not rigorously always a lattice,
causal sets have been included here in order to also exemplify, in
as close a fashion as possible, the use of a lattice of the second
type in the construction of an alternative structure for
spacetime.\footnote{Causal sets can also be classified as
graph-theoretical, considering that they can be thought of as
consisting of a locally-finite, loopless, directed graph without
circuits, in the sense that it is not possible to come back to a
starting vertex while following the allowed directions in it.}

\subsubsection{Simplicial quantum gravity by Lehto et al.}
Lehto (1988), and Lehto, Nielsen, and Ninomiya (1986a,b) attempted
the construction of a pregeometric scheme in which it was
conjectured that spacetime possesses a deeper, pregeometric
structure described by three dynamical variables, namely, an
abstract simplicial complex ASC, its number of vertices $n$, and a real-valued field $\ell$ associated with every one-simplex (pair of vertices) $Y$. It is significant that, up to this point, the approach has a chance of being truly pregeometric since the fields $\ell$ are not (yet) to be understood as lengths of any sort, and the vertices could also be thought as (or called) elements or abstract objects (cf.\ \emph{abstract} simplicial complex). However, the introduction of geometric concepts follows
immediately.

Next, an abstract simplicial complex is set into a unique
correspondence with a geometric simplicial complex GSC, a scheme
of vertices and geometric simplices. By piecing together these
geometric simplices, a piecewise linear space can be built so
that, by construction, it admits a triangulation. Furthermore, the
crucial step is now taken to interpret the above field $\ell$ as
the link \emph{distance} of the piecewise linear manifold. The
conditions are thus set for the introduction of a Regge calculus
lattice with metric $g_{\mu\nu}^{\mathrm{RC}}$ given by the
previously defined link lengths.

Traditionally, the primary idea of Regge calculus is to provide a
piecewise linear approximation to the spacetime manifolds of
general relativity by means of the gluing together of
four-dimensional simplices, with curvature concentrated only in
their two-dimensional faces. In this approach, however, the
approximation goes in the opposite direction, since simplicial
gravity is now seen as more fundamental, having the smooth
manifolds of general relativity as a large-scale limit. Moreover,
whereas the basic concept of the Regge calculus consists in the
link lengths, which once given are enough to determine the
geometry of the lattice structure, Lehto et al.\ also introduced
the number of vertices $n$ as a dynamical variable. This conferred
on the lattice vertices a quality resembling that of a free gas,
in the sense that any pair of vertices can with high probability
have any mutual distance, helping thus to avoid the rise of
long-range correlations typical of fixed-vertex lattices.

A quantization of the Regge calculus lattice is subsequently
introduced by means of the Euclidean path-integral formalism. A
pregeometric partition function $Z$ summing over $\mathrm{ASC}$,
fields $\ell$, and vertices $n$ is first proposed as a formal
abstraction, to be geometrically realized by a corresponding Regge
calculus partition function $Z^\mathrm{RC}$ summing over piecewise
linear manifolds and link lengths $\ell$. Indeed, quantization can
be viewed as one of the reasons why the notion of length must be
introduced in order to move forwards.

Two reasons can be offered for the choice of the starting point of
this framework. On the one hand, being concerned with the
production of a genuine pregeometry, an abstract simplicial
complex was chosen on the grounds that, as such, it is abstract
enough to be free from geometric notions intrinsic to it. On the
other hand, an abstract simplicial complex with a variable set of
elements was seen fit to provide an appropriate setting in
which to study further and from a different point of view the
character of diffeomorphism invariance, itself a crucial element
for a possible richer understanding of the physical existence of
spacetime. This is because the requirement of diffeomorphism
symmetry of the functional integral measure $\prod
\mathrm{d}\ell\exp(-S_\mathrm{pg})$, where $S_\mathrm{pg}$ is a
Euclidean pregeometric action, with respect to a displacement of
vertices results in a free-gas behaviour of the latter as elements
of a geometric lattice.

In one sense, this work is no exception among all those analyzed
in this section. As it was seen, geometric concepts such as link
length had to be assumed to realize or materialize the abstract
simplicial complex as a geometric object, a Regge calculus
lattice, bringing it thus closer to the more familiar geometric
world and rendering it fit for quantization. However, despite the
assumption of geometry, we would nevertheless like to lay emphasis
on a distinctive feature of this work---it recognizes clearly
pregeometric and geometric realms and, even though it cannot do
without the latter, it keeps them clearly distinguished from one
another rather than, as is more common, soon losing sight of all
differences between them.

\subsubsection{Causal sets by Bombelli et al.}
Bombelli, Lee, Meyer, and Sorkin (1987) proposed that at the
smallest scales spacetime is a causal set: a locally finite set of
elements endowed with a partial order corresponding to the
macroscopic relation that defines past and future. This
partial-order relation is required to have transitivity and
irreflexivity properties of causality between point-events.

In this framework, causal order is viewed as prior to metric and
not the other way around; in more detail, the differential
structure and the conformal metric of a manifold are derived from
a causal order. Because only the conformal metric (it has no
associated measure of length) can be obtained in this way, the
transition is made to a non-continuous spacetime consisting of a
finite but large number of ordered elements, a causal set $C$. In
such a space, size can be measured by counting. The transition to
the classical limit is expressed as the possibility of a faithful
embedding of $C$ in $M$, or of a coarse grained version $C'$ of
$C$, arising from the assumption that small-scale fluctuations
would render $C$ non-embeddable.

What considerations have led to the choice of a causal set? In
this respect, Sorkin explained:
\begin{quote}
The insight underlying these proposals is that, in passing from
the continuous to the discrete, one actually \emph{gains} certain
information, because ``volume'' can now be assessed (as Riemann
said) \emph{by counting}; and with both order \emph{and} volume
information present, we have enough to recover geometry. (Sorkin,
2005, p.~5\footnote{Page number refers to the online preprint
gr-qc/0309009 v1.})
\end{quote}
This is a valid recognition resting on the fact that, given that
general relativity's local metric is fixed by causal structure and
conformal factor, a discrete causal set will be able to reproduce
these features and also provide a way for the continuous manifold
to emerge. However, we note that Sorkin is not here following a
principle in the sense of Section \ref{PEST} that naturally leads
him to the conclusion that a causal set structure underlies
spacetime. Furthermore, volume is already a geometric concept, so
that ``to recover geometry'' above must mean ``to recover a
continuous, metric spacetime,'' not ``geometry'' in the stricter
sense of the word.

Finally, the use of further geometric concepts in the formulation
of this picture becomes clear in (Sorkin, 1991). The
correspondence principle between a manifold $(M,\mathbf{g})$ and a
causal set $C$ from which it is said to emerge, for example,
reads:
\begin{quote}
The manifold $(M,g)$ ``emerges from'' the causal set $C$ iff $C$
``could have come from sprinkling points into $M$ at unit density,
and endowing the sprinkled points with the order they inherit from
the light-cone structure of $g$.'' (Sorkin, 1991, p.~156)
\end{quote}
The above-mentioned unit density means that there exists a
\emph{fundamental unit of volume}, expected to be the Planck
volume, \mbox{$10^{-105}$ m$^3$}. Furthermore, Sorkin also
considered the \emph{distance} between two causal set elements $x$
and $y$ as the number of elements in the longest chain joining them  (p.~17), thus turning the ordering
relations into lines and, possibly, $x$ and $y$ into points. Thus, if a causal set is
to be seen as a pregeometric framework, due to its geometric
assumptions, it makes again a questionable case of it. However, we
do not know whether its proponents have staked any such claim.

\subsection{Number-theoretical pictures}
Some investigators have identified the key to an advancement in
the problems of spacetime structure as lying in the number fields
used in current theories or to be used in future ones. Butterfield
and Isham (2000, pp.~84--85), for example, arrived at the
conclusion that the use of real and complex numbers in quantum
mechanics presupposes that space is a continuum. According to this
view, standard quantum mechanics could not be used in the
construction of any theory that attempts to go beyond such a
characterization of space.

\subsubsection{Hill's discrete spacetime}
Hill (1955) assumed that the structure of space is determined by
its allowed symmetry transformations. Instead of requiring invariance of his theory under the full continuous group of Lorentz transformations $L$, he selected the subgroup of $L$ named $L_r$ in which translations and Lorentz transformations have rational, rather than real, coefficients. This subgroup is a group with respect to successive transformations and, in addition, it is dense in the full group $L$. From the latter property it follows that the deviation of Hill's picture from the continuous treatment of special relativity leaves very little room for its disproof, as he himself stated.

As the energy-momentum space is first discretized by restricting
the values of energy and momentum variables to a certain set of
rational numbers, quantum-mechanical wave functions become a
special case of almost periodical functions, having space and time as continuous variables. When the change to spacetime points with only rational coordinates is made, the desired invariance of the background is realized, but the essence of the wave functions is obscure since the energy-momentum space lacks a unique
interpretation.

Thus, guided by the Lorentz invariance of special relativity and
the concept of wave function in quantum mechanics, Hill presented
a picture in which some degree of success is achieved as regards
the unification of ideas from the two theories; compared with some
earlier lattice models, the problems of breaking the Lorentz
invariance and of having an impossibly large minimum velocity
(such as in Schild's) are solved. However, as Hill admitted, the
implications of his model---especially experimentally testable
ones---are not well-known.

Finally, since Hill assumed the structure of traditional quantum
mechanics and of special relativity, the role of geometry is
ubiquitous in his theory. However, since he did not endeavour to go further and apply his ideas to the construction of some form or other of pregeometry, his work does not attempt to go a great deal beyond today's received theories; therefore, the usual criticism made in this section as to the surreptitious use of geometric notions does not apply.

\subsubsection{Rational-number spacetime by Horzela et al.}
Horzela, Kapu\'{s}cik, Kempczy\'{n}ski, and Uzes (1992) criticized
those discrete representations of spacetime that assume an
elementary length, and which may furthermore violate relativistic
invariance, on the grounds that the former is experimentally not
observed and the latter is, in fact, experimentally
verified. They proposed to start their analysis by studying the
actual experimental method used to measure spacetime coordinates,
the radio-location method.

Their basic claim is that the measured values of the coordinates
$t=(1/2)(t_1+t_2)$ and $x=(c/2)(t_2-t_1)$ of an event in any
system of reference are always rational (rather than real)
numbers, since both the time of emission $t_1$ and reception $t_2$
of the radio signal are crude, straightforward measurements made
with a clock.\footnote{For this idea to work, it is also essential that the value $c$ of the speed of light be understood as the result of a ``crude'' measurement.} Subsequently, they showed that the property of being
rational is preserved when calculating the coordinates of an event
in another system, as well as in the calculation of the relative
velocity of this system. Therefore, they maintained that Lorentz
invariance holds in a spacetime pictured in terms of
rational-number coordinates. In addition, since the rational
numbers are dense in the real numbers, such a spacetime frees
itself from any notion of elementary distance. Thus, if spacetime
must be described in a discrete manner, the rational number field
is then seen as the key to a better understanding of it.

While Horzela et al. suggested that this line of investigation
should be furthered by means of algebraic mathematical methods,
they admitted not knowing at the time of their writing of any
guiding physical principle to do this. As a consequence, since
they do not propose to go as far as building a more complete
spacetime picture based on their preliminary findings, the usual
criticisms being made in this section do not apply. Further
physical guiding principles are lacking, but then this has been
clearly expressed and action was taken in accord with it;
geometric ideas are of course present, but then no real attempt to
transcend the framework of relativity was made after all.

\subsubsection{Volovich's number theory}
Volovich (1987) argued that, since at the Planck scales the usual
notion of spacetime is suspected to lose its meaning, the building
blocks of the universe cannot be particles or fields or strings
for these are defined on such a background; in their place, he
proposed that the numbers be considered as the basic entities.
Moreover, Volovich suggested that for lengths less than the Planck
length, the Archimedean axiom\footnote{The Archimedean axiom
states that any given segment of a straight line can be eventually
surpassed by adding arbitrarily small segments of the same line.}
may not hold. But why---he asked---should one construct a
non-Archimedean geometry over the field of real numbers and not
over some other field? In this he considered the field of rational
numbers $\mathbb{Q}$ and a finite Galois field, since these two
are contained as subfields of any other field.

The question above is inspired in Volovich's proposed general
principle that all physical parameters undergo quantum
fluctuations, including the number field on which a theory is
developed (p.~14). He maintained that this means that the usual
field of real numbers is apparently only realized with some
probability but that, in principle, there also exists the
probability of the occurrence of any other field. His programme
therefore consists in reducing ``all physics'' to geometry over
arbitrary number fields. Volovich furthermore speculated that the
fundamental physical laws should be invariant under a change of
the number field, in analogy with Einstein's principle of general
invariance (p.~14).

In his view, ``number theory and the corresponding branches of
algebraic geometry are nothing else than the ultimate and unified
physical theory'' (p.~15). We judge such a statement not only too
quick but also as staking a claim about a physical theory that is
more definitive than any such theory could ever be.

Furthermore, it is difficult to see what principle led Volovich to
propose this programme; it appears that only analogies have guided
his choices. Finally, the role of geometry in this framework is
crystal-clear: instead of attempting to reduce ``all physics'' to
Archimedean geometry over the real numbers, it is suggested that
one should rather attempt to reduce ``all physics'' to
non-Archimedean \emph{geometry} over an arbitrary number field.
For this reason, although Volovich did not so claim, his scheme
could, again, hardly be considered a true instance of pregeometry.

\subsection{Relational or process-based pictures}
Having found inspiration in Leibniz's relational conception of
space, some researchers have proposed that spacetime is not a
thing itself but a resultant of a complex of relations among basic
things. Others, on a more philosophical strand, have found
inspiration in Heraclitus's principle that ``all is flux'' and
have put forward pregeometric pictures in which, roughly speaking,
processes are considered to be more fundamental than things. This
view is sometimes linked with the currently much-disseminated idea
of information, especially in its quantum-mechanical form.

\subsubsection{Axiomatic pregeometry by Perez Bergliaffa et al.}
Perez Bergliaffa, Romero, and Vucetich (1998) presented an
axiomatic framework that, via rules of correspondence, is to serve
as a ``pregeometry of space-time.'' This framework assumes the
objective existence of basic physical entities called things and
sees spacetime not as a thing itself but as a resultant of
relations among those entities.

In order to derive from this more basic substratum the topological
and metric properties of Minkowskian spacetime, a long list of
concepts and axioms is presented. But for the said general
requirements of objectivity and relationality, these ``ontological
presuppositions'' (p.~2283) appear arbitrary rather than stemming
from some physical principle concerning spacetime.

Among these concepts, we would like to centre our attention on
something called ontic space $E_{\mathrm{o}}$. Although its exact
meaning cannot be conveyed without reviewing a great deal of the
contents of the article in question, it could be intuitively taken
to be a form of space connected with the basic entities, which is
prior (in axiomatic development) to the more familiar geometric
space $E_{\mathrm{G}}$ of physics---Minkowskian space, in this
case. Perez Bergliaffa et al.\ proved (pp.~2290--2291) that the
ontic space $E_{\mathrm{o}}$ is metrizable and gave an explicit
metric $d(x,y)$ for it, a \emph{distance} between things, turning,
again, relations into lines and, possibly, things into points.
Finally, by means of an isomorphism between $E_{\mathrm{o}}$ and a
subspace dense in a complete space, the geometric space
$E_{\mathrm{G}}$ is obtained as this complete space itself, while
at the same time it inherits the metric of $E_{\mathrm{o}}$.

One notices once again the stealthy introduction of a metric for
pregeometry, with the ensuing geometric objectification of earlier
abstract concepts, in order to derive from it further geometric
notions. This procedure must be criticized in view that this
approach explicitly claims to be an instance of pregeometry.

\subsubsection{Cahill and Klinger's bootstrap universe}
Cahill and Klinger (1997) proposed to give an account of what they
rather questionably called ``the `ultimate' modelling of reality''
(p.~2). This, they claimed, could only be done by means of
pregeometric concepts that do not assume any notions of things but
only those of processes since, according to them, the former
notions (but not the latter) suffer from the problem of their
always being capable of further explanation in terms of new
things. One must wonder for what mysterious reason certain
processes cannot be explained in terms of others.

As an embodiment of the concepts above, these authors used a
``bootstrapped self-referential system'' (p.~3) characterized by a
certain iterative map that relies on the notion of relational
information $B_{ij}$ between monads $i$ and $j$. In order to avoid
the already rejected idea of things, these monads are said to have
no independent meaning but for the one they acquire via the
relational information $B_{ij}$.

Due to reasons not to be reviewed here, the iterative map will
allow the persistence of large $B_{ij}$. These, it is argued, will
give rise to a tree-graph---with monads as nodes and $B_{ij}$ as
links---as the most probable structure and as seen from the
perspective of monad $i$. Subsequently, so as to obtain (by means
of probabilistic mathematical tools) a persistent background
structure to be associated with some form of three-dimensional
space, a \emph{distance} between any two monads is defined as the
smallest number of links connecting them in the graph.

``This emergent 3-space,'' Cahill and Klinger argued, ``\ldots
does not arise within any \emph{a priori} geometrical background
structure'' (p.~6). Such a contention is clearly mistaken for a
graph with its defined idea of distance is certainly a geometric
background for the reasons we have pointed out earlier. Thus, yet
another pregeometric scheme falls prey to geometry.

\subsection{Quantum-cosmological pictures} \label{QCP}
Quantum-cosmological approaches tackle the problem of spacetime
structure from the perspective of the universe as a whole and seen
as a necessarily closed quantum system. For them, the problem of
the constitution of spacetime is bound up with the problem of the
constitution of the cosmos, customarily dealing as well with the
problem of how they interdependently originated.

\subsubsection{Eakins and Jaroszkiewicz's quantum universe} \label{EJ}
Jaroszkiewicz (2001) and Eakins and Jaroszkiewicz (2002, 2003)
presented a theoretical picture for the quantum structure and
running of the universe. Its first class of basic elements are
event states; these may or may not be factored out as yet more
fundamental event states, depending on whether they are direct
products (classicity) of yet more elementary event states
$|\psi\rangle$, or whether they are entangled (non-separability)
event states $|\Psi\rangle$. The second class of basic elements
are the tests acting on the event states. These tests $\Sigma$,
represented by Hermitian operators $\hat{\Sigma}$, provide the
topological relationships between them, ultimately endowing the
structure of states with an evolution (irreversible acquisition of
information) via ``state collapse'' and a quantum arrow of time. A
third component is a certain information content of the
universe---sometimes, although not necessarily, represented by the
semi-classical observer---at each given step in its evolution
(stage), which, together with the present state of the universe,
determines the tests which are to follow. Thus, the universe is a
self-testing machine, a quantum automaton, in which the
traditional quantum-mechanical observer is explained but totally
dispensed with.

In (Eakins \& Jaroszkiewicz, 2003, p.~1), these authors found
a connection between causal set theory and their own framework. In
the latter, two different causal set structures arise: one in
connection with the separation and entanglement of states, seen to
be related to the transmission of quantum-mechanical information
which does not respect Einstein's locality; the other in
connection with the separation and entanglement of operators, seen
to be related to the transmission of classical information.

Eakins and Jaroszkiewicz (2002, p.~5), like Requardt but after
their own fashion, also attempted to understand the present
structure of spacetime as a consequence of primordial events,
which in this case can be traced back to what they call the
``quantum big bang.''\footnote{In this case, the emergence of spacetime is more sophisticated than in Requardt's case, since here it happens in the non-parameter, intrinsic time given by state vector collapse.} Hence the classification of this approach.

The subject of pregeometry as pioneered by Wheeler is also
touched upon (Eakins \& Jaroszkiewicz, 2003, p.~2). The view that
the characteristic feature of pregeometric approaches consists in
``avoiding any assumption of a pre-existing manifold'' (p.~2) is
also stated there, although it is nowhere mentioned that the
notion of distance, according to Wheeler, should be avoided as
well. In any case, these authors explained that their task is to
reconcile pregeometric, bottom-up approaches to quantum gravity
with other holistic, top-bottom ones as in quantum cosmology.

Since the traditional machinery of quantum mechanics is assumed in
the approach, so is therefore geometry (see Section \ref{GQMec}).
Again, this fact would undermine the status of this programme as a genuine representative of pregeometry. \mbox{(See Table 1.)}

Of all the above approaches, this one is perhaps the most
physically inspired one in the sense that it gives clear
explanations about the nature of its building blocks and puts
forward judicious physical principles and mechanisms---as
contrasted to purely mathematically inspired ones---by means of
which the traditional notions of spacetime and quantum
non-locality may emerge. 

\begin{table} \centering
\begin{tabular}{|l|c|c|} 
\hline
\textbf{Author} & \textbf{Geometric objects} & \textbf{Geometric magnitudes}\\
\hline \hline 
Dadi\'{c} \& Pisk & vertices, edges & overlap, length, distance \\ \hline
Antonsen & idem & length, distance \\ \hline
Requardt & idem & idem \\ \hline
Lehto et al. & idem & idem \\ \hline
Bombelli et al. & idem & length, distance, volume \\ \hline
Hill & as in sp.~rel. \& q.~mech. &  as in sp.~rel. \& q.~mech. \\ \hline
Horzela et al. & as in sp.~rel. &  as in sp.~rel.\\ \hline
Volovich & as in mod.~phys. &   as in mod.~phys. \\ \hline
Perez Bergliaffa et al. & points, lines &  length, distance \\ \hline
Cahill \& Klinger & idem &  idem \\ \hline
Eakins \& Jaroszkiewicz & as in q.~mech. &  as in q.~mech. \\ \hline
Nagels & points, lines & length, distance \\ \hline
Stuckey \& Silberstein & idem & idem \\ \hline
\end{tabular}
\caption{Summary of the use of geometric concepts in the pregeometric approaches analyzed, including those of Section \ref{bucket}. Abbreviations: Special relativity (sp.~rel.), quantum mechanics (q.~mech.), modern physics (mod.~phys.).}
\end{table}

\subsection{The inexorability of geometric understanding?}
Finally, we would like to take note of certain views expressed by
Anandan, as they constitute the absolute epitome of pregeometry's
mode of working described so far. Anandan is led to suggest
\begin{quote}
a philosophical principle which may be schematically expressed as
\begin{center} Ontology = Geometry = Physics.\end{center}
The last equality has not been achieved yet by physicists because
we do not have a quantum gravity. But it is here proposed as a
philosophical principle which should ultimately be satisfied by a
physical theory. (Anandan, 1997, p.~51)
\end{quote}
Anandan appears to have taken the geometric essence of the present
description of spacetime as a ground to argue ahead that,
correspondingly, any successful future description of it must of
necessity be geometric. Such is his conviction that he raises this
idea to the status of a principle that any physical theory should
``ultimately'' follow. But does it ensue from the fact that
physics has so far described Nature by means of geometry that it
will continue to do so in the future? Physics \emph{must} describe
Nature geometrically come what may---\emph{whence such an
inevitability?} Moreover, it is not clear in what way the first
part of the equality has, as implicitly stated, already been
achieved. Is this a suggestion that reality itself is geometric?
It is hard to know what to make of such a cryptic notion.

Thus, one witnesses here an overstated version of the cases
analyzed through this section. If in the latter the use of
geometry was, so to speak, accidental or non-intentional, in
Anandan's statement one sees the explicit claim materialized that
geometry \emph{must} be the mode of description of physical
science, in particular, concerning a solution to the problem of
the structure of spacetime. No stronger grip of geometry on
physics could possibly be conceived. This view clashes head-on
with our own as to what the role of geometry could be in these
matters.

\subsection{Appraisal}
The connection with the first part of this article now comes to
the fore. From the mists of antiquity to the forefront of
twenty-first century theoretical physics, \emph{geometry remains
the enduring, irreplaceable tool by means of which humans portray
the world}. Ironically enough, even in a field that sets itself
the task of explaining the origin of geometry in physics and names
itself pregeometry, one is to find the most explicit displays of
geometric understanding. Some reasons suspected to have a bearing
on this peculiar state of affairs will be offered in the following
section.

\section{Beyond geometry} \label{BG}
It is the cherished realm of geometry that, some have proposed,
might need to be transcended in the search for a novel theory of
spacetime structure---or more profoundly still, in the search for
a deeper layer of the nature of things. In this section, some of
the views espoused by Clifford in 1875, by Eddington in 1920, and
by Wheeler in the decades between 1960 and 1980 will be explored
as examples of this sort of proposal. In Wheeler's case, the
reason for such a requirement will be seen to stem directly from
the general views of Section \ref{PEST}, namely, that progress in
the physical world picture comes after providing explanations of
Nature of a fundamentally different kind than those currently in
force. Clifford and Eddington do not make an explicit case for
this point, but their ideas can be seen to rest on similar
foundations.

\subsection{Clifford's elements of feeling} \label{CEF}
Views that seem to advocate the provisional character of geometric
explanations in physical theories date at least as far back as
1875. For then, not only did Clifford express his better-known
belief that matter and its motion are in fact nothing but the
curvature of space---and thus reducible to geometry---but also
lesser-known ideas about matter and its motion---and so perhaps,
indirectly, geometry---being, in turn, only a partial aspect of
``the complex thing we call feeling.''

Within the context of an investigation to which we do not wish to
commit ourselves, but from which we would nevertheless like to
rescue an interesting idea, Clifford (1886, pp.~172--173) asked
about the existence of something that is not part of the
``material or phenomenal world'' such as matter and its motion,
but that is its ``non-phenomenal counterpart.'' After a linguistic
reinterpretation of Clifford, this issue can be understood to be
connected with the search for a new brand of theoretical entities,
on a deeper level than the presently accepted ones, and on the
basis of which to provide a new explication of Nature. In this
regard, Clifford wrote:
\begin{quote}
The answer to this question is to be found in the theory of
sensation; which tells us not merely that there is a
non-phenomenal world, but also in some measure what it is made of.
Namely, the reality corresponding to our perception of the motion
of matter is an element of the complex thing we call feeling. What
we might perceive as a plexus of nerve-disturbances is really in
itself a feeling; and the succession of feelings which constitutes
a man's consciousness is the reality which produces in our minds
the perception of the motions of his brain. These elements of
feeling have relations of nextness or contiguity in space, which
are exemplified by the sight-perceptions of contiguous points; and
relations of succession in time which are exemplified by all
perceptions. Out of these two relations the future theorist has to
build up the world as best he may. (Clifford, 1886, p.~173)
\end{quote}

Allowing for 129 years of history, it could be said that Clifford
pursued after this particular fashion the conviction that matter
and geometry do not lie at the bottom of things, but that, on the
contrary, they are themselves an aspect of deeper entities---for
him, feelings---and their non-geometric relations, contiguity and
succession. It is in these objects beyond matter and geometry
that, Clifford believed, one can catch a glimpse of a deeper layer
of the constitution of the world.

\subsection{Eddington's nature of things}
Forty-five years later, Eddington (1920, pp.~180--201) expressed
similar views as to what he called the nature of things. In
analyzing general relativity, he identified its basic elements as
the point-events, and the interval as an elementary relation
between them. As to the nature of the latter, Eddington wrote:
\begin{quote}
Its [the interval's] geometrical properties\ldots can only
represent one aspect of the relation. It may have other aspects
associated with features of the world outside the scope of
physics. But in physics we are concerned not with the nature of
the relation but with the number assigned to express its
intensity; and this suggests a graphical representation, leading
to a geometrical theory of the world of physics. (Eddington, 1920,
p.~187)
\end{quote}
Along these lines (which accord with our views of Section
\ref{Geometry}), he suggested that the \emph{individual} intervals
between point-events probably escape today's scales and clocks,
these being too rudimentary to capture them. As a consequence, in
general relativity one only deals with macroscopic values composed
out of many individual intervals. ``Perhaps,'' he ventured further
in an allusion to transcending geometric magnitudes,
\begin{quote}
even the primitive interval is not quantitative, but simply 1 for
certain pairs of point-events and 0 for others. The formula given
[$\mathrm{d}s^2=g_{\mu\nu}\mathrm{d}x^\mu \mathrm{d}x^\nu$] is
just an average summary which suffices for our coarse methods of
investigation, and holds true only statistically. (Eddington,
1920, p.~188)
\end{quote}
One must not be confused by the association of the numbers 1 and 0
to the primitive intervals and think that such numbers represent
their lengths. In the passage, the \emph{non-quantitativeness} of
the primitive intervals is clearly stated. What the 1 and 0
represent is merely whether the intervals exist or not, any other
general designation having been equally satisfactory for the
purpose. Eddington's insight into the need to transcend geometric
notions is reinforced by his further remarks: ``[W]e can scarcely
hope to build up a theory of the nature of things if we take a
scale and a clock as the simplest unanalysable concepts''
(p.~191). It is nothing short of remarkable to find expressed such
deep views, on the one hand, inspired in the general theory of
relativity, while on the other, in a sense contrary to its spirit,
only five years after the publication of the latter. We note,
however, that Eddington's avowal only seems to be for the
overthrow of geometric magnitudes but not of geometric objects,
since he does not seem to find fault with the concepts of
point-event and interval.

Eddington also stressed the role of the human mind in the
construction of physical theories. He argued that, out of the
above primitive intervals (taken as elements of reality and not of
a theory of it) between point-events, a vast number of more
complicated qualities can arise; as a matter of fact, however,
only certain qualities of all the possible ones do arise. Which
qualities are to become apparent (in one's theories) and which not
depends, according to Eddington, on which aspects of these
elementary constituents of the world (and not of a theory of it)
the mind singles out for recognition. ``Mind filters out matter
from the meaningless jumble of qualities,'' he said,
\begin{quote}
as the prism filters out the colours of the rainbow from the
chaotic pulsations of white light. Mind exalts the permanent and
ignores the transitory\ldots\ Is it too much to say that mind's
search for permanence has created the world of physics? So that
the world we perceive around us could scarcely have been other
than it is? (Eddington, 1920, p.~198)
\end{quote}
At this point, a clarification is in order in connection with the
parenthetical additions above and with Eddington's remarks. He
talked about the mind creating the world of physics, but one finds
such a thing a daring exploit, for how could the mind create the
physical world? Or to put it in the words of Devitt and Sterelny
(1987), ``how could we, literally, have made the stars?'' (p.~200). Rather, what the mind creates is \emph{theories of the world} and into these inventions it puts, naturally, all its partialities and prejudices.

This being said, one feels tempted to add to Eddington's
observations, as related ideas promptly suggest themselves. Matter
is but one of a whole range of entities that physical theories
find to be conserved. Paraphrasing Eddington, one could therefore
ask: is it too much to say that mind's search for conservation
laws and symmetries has created the current theories of the
physical world? What is more, after our analysis of geometry and pregeometry, along the
same lines one could furthermore ask: is it too much to say that
\emph{mind's geometric instinct} has created all the current
theories of the physical world?

\subsection{Wheeler's pregeometry} \label{WP}
Wheeler (1964, 1980), Misner et al.\ (1973), and Patton and
Wheeler (1975) expressed pioneering ideas on what they called
pregeometry already four decades ago. Since we understand Wheeler
to be by far the main contributor to this idea, the kind of
pregeometry to be analyzed in this section will be correspondingly
called Wheeler's pregeometry. In order to avoid confusion, it must
be remarked here yet again---in addition to the early comments of
Section \ref{PG}---that Wheeler's pregeometry is not really what
later on came to be known by the same name, i.e.\ pre-manifold
physics, but the more radical stance of going beyond geometry in a
proper sense.

His basic demand amounts to the rejection of geometric concepts in
order to explain geometric structure. Indeed, Wheeler (1980)
advocated ``a concept of pregeometry that breaks loose at the
start from all mention of geometry and distance;'' he was wary of
schemes in which ``too much geometric structure is presupposed to
lead to a believable theory of geometric structure;'' and he
clearly recognized that ``to admit distance at all is to give up
on the search for pregeometry'' (pp.~3--4). This means that one
surprisingly finds the very inventor of pregeometry generally
invalidating all the schemes analyzed in Section \ref{PG}, as far
as they are to be an expression of his original pregeometry.

What are the grounds for these strong pronouncements of Wheeler's?
Firstly, he envisaged (Patton \& Wheeler, 1975, p.~547; Misner et
al., 1973, p.~1201), among other things, spacetime collapse in the
form of the big bang, a possible big crunch, black holes, and a
supposed foam-like structure of spacetime on small scales, as
indicators that spacetime cannot be a continuous manifold, since
it has the ability to become singular. Confronted thus by the need
to rethink spacetime, he posed the question:
\begin{quote}
If the elastic medium is built out of electrons and nuclei and
nothing more, if cloth is built out of thread and nothing more, we
are led to ask out of what ``pregeometry'' the geometry of space
and spacetime are built. (Wheeler, 1980, p.~1)
\end{quote}
The reason why Wheeler seeks a ``pregeometric building material''
(Misner et al., 1973, p.~1203) in order to account for the
spacetime continuum is that he believes that genuine explanations
about the nature of something do not come about by explicating a
concept in terms of similar ones, but by reducing it to a
different, more basic kind of object. This is evidenced in the
passage: ``And must not this something, this `pregeometry,' be as
far removed from geometry as the quantum mechanics of electrons is
far removed from elasticity?'' (p.~1207)

Having stated Wheeler's motives for going beyond geometry, what
were his proposals to implement this programme? His attempts
include pregeometry as a Borel set or bucket of dust (see Section \ref{bucket}),
pregeometry as binary-choice logic, and pregeometry as a
self-referential universe.\footnote{See (Demaret, Heller, \&
Lambert, 1997, pp.~157--161) for an independent review of
Wheeler's pregeometry.}

An early attempt at pregeometry based on binary choice was
Wheeler's ``sewing machine,'' into which rings or loops of space
were fed and connected or left unconnected by the machine
according to encoded binary information (yes or no) for each
possible pair of rings (Misner et al., 1973, pp.~1209--1210).
Wheeler considered the coded instructions to be given from without
either deterministically or probabilistically, and then asked
whether fabrics of different dimensionality would arise, and
whether there would be a higher weight for any one dimensionality
to appear. We note that, as a design for pregeometry, this idea
succeeds in staying clear of geometric magnitudes but fails, at
least superficially, to avoid geometric objects---Wheeler's loops
of space have no size and yet are \emph{rings}. We say this
failure is superficial because these rings or loops, so long as
they have no size, could as well be called abstract elements.
If nothing else, the mention of rings is witness to the fact that
geometric modes of thought are always lurking in the background. 

A subsequent idea still based on binary-choice logic was pregeometry as the calculus of propositions (Misner et al., 1973,
pp.~1209, 1211--1212). This conception was much less picturesque
and rather more abstract. Wheeler exploited the isomorphism
between the truth-values of a proposition and the state of a
switching circuit to toy with the idea that `` `physics'
automatically emerges from the statistics of very long
propositions, and very many propositions,'' (Patton \& Wheeler,
1975, p.~598) in thermodynamical analogy. In this reference and in (Wheeler, 1980), however, the expectation that pure mathematical logic \emph{alone} could have anything whatsoever to do with providing a foundation for physics was sensibly acknowledged. In this case, we note that, perhaps due to its abstractness and total divorcement with physics, this conception of pregeometry is truly free of all geometry in a strict sense. For all this scheme was worth, it is at least rewarding to find it has this feature.

Later on, inspired in self-referential propositions, Wheeler
conceived of the idea of pregeometry as a self-referential
universe. As with the previous approaches, Wheeler (1980) admitted
to having no more than a vision of how this understanding of
geometry in terms of (his) pregeometry might come about. Two basic
ingredients in this vision appear to be (i) events consisting of a
primitive form of quantum principle: investigate Nature and create
reality in so doing (see below) and, more intelligibly, (ii)
stochastic processes among these events producing form and
dependability out of randomness. His vision reads thus:
\begin{quote}
(1) Law without law with no before before the big bang and no
after after collapse. The universe and the laws that guide it
could not have existed from everlasting to everlasting. Law must
have come into being\ldots\ Moreover, there could have been no
message engraved in advance on a tablet of stone to tell them how
to come into being. They had to come into being in a
higgledy-piggledy way, as the order of genera and species came
into being by the blind accidents of billions upon billions of
mutations, and as the second law of thermodynamics with all its
dependability and precision comes into being out of the blind
accidents of motion of molecules who would have laughed at the
second law if they had ever heard of it. (2) Individual events.
Events beyond law. Events so numerous and so uncoordinated that
flaunting their freedom from formula, they yet fabricate firm
form. (3) These events, not of some new kind, but the elementary
act of question to nature and a probability guided answer given by
nature, the familiar everyday elementary quantum act of
observer-participancy. (4) Billions upon billions of such acts
giving rise, via an overpowering statistics, to the regularities
of physical law and to the appearance of continuous spacetime.
(Wheeler, 1980, pp.~5--6)
\end{quote}

Earlier, Patton and Wheeler (1975, pp.~556--562) had elaborated
further on this. They considered two alternatives to the problem
of spacetime collapse, namely: (a) taking into account quantum
mechanics, the event of universal collapse can be viewed as a
``probabilistic scattering in superspace.'' In this way, the
universe is reprocessed at each cycle, but the spacetime manifold
retains its fundamental mode of description in physics. This
alternative was not favoured by these authors, but instead (b) considering as
an overarching guiding principle that the universe must have a way
to come into being (cosmogony), they suggested that such a
requirement can be fulfilled by a ``quantum principle'' of
``observer-participator.''\footnote{This principle is tantamount
to the first ingredient (i) above, and perhaps helps to clarify
it.} According to it, the universe goes through one cycle only,
but only if in it a ``communicating community'' arises that can
give meaning to it. This universe is thus self-referential, which
was the standpoint favoured by these authors.

We cannot agree with the latter proposal. For although it is
correct to say that only a communicating community of beings---
which furthermore engage in scientific activities---can give
meaning to the displays of Nature, it is beyond any reasonable
doubt that the universe exists and moves on independently of any
consciousness that it might give rise to. Finally, as far as
pregeometric ideas are concerned, this vision of self-referential
cosmology succeeds in eschewing all kinds of geometric concepts
again due to its being conceived in such broad and general terms.
Regardless of its worth and in the general context of paradoxical
geometric pregeometry, let this be a welcome feature of Wheeler's
effort.

Regardless of Wheeler's actual implementations of his pregeometry
and any criticisms thereof, the imperative demand that the
foundations of such a theory should be completely free of
geometric concepts remains unaltered. The fact that so many
(including, to the smallest extent, Wheeler himself) have misconstrued
his idea of a programme for pregeometry is perhaps a tell-tale
sign of something beyond simple carelessness in the reading of his
views. Indeed, the large-scale misinterpretation here documented
may just as well stem from the lack of all background for thought
that one encounters as soon as one attempts to dispense with
geometry. Physics knows of no other mode of working than geometry,
and thus physicists have reinterpreted Wheeler's revolutionary new
idea in the only way that made sense to them---\emph{setting out
to find geometry in pregeometry.}

\subsubsection{More buckets of dust} \label{bucket}
As two final claimants to pregeometry to be analyzed, we turn to
the works of Nagels and of Stuckey and Silberstein in the light of
their close connection with Wheeler. In these works, their authors
independently attempt again to build ``space as a bucket of dust''
after Wheeler's (1964, pp.~495--499) earlier effort. This earlier
effort, clearly connected with the above-mentioned ``sewing
machine'' in character and purpose, consisted in starting with a
Borel set of points without any relations to each other whatsoever
and assembling them into structures of different dimensionality on
the basis of different quantum-mechanical probability amplitudes
attributed to the relations of nearest neighbour between the
points. Wheeler dismissed his own trial because, among other
reasons, the same quantum-mechanical principles used to define
adjacency rendered the idea untenable since points that had once
been neighbours would remain correlated after departing from each
other (p.~498). We observe once more that this idea, in its use of
the concept of points (cf.\ earlier rings), is not free from
geometric objects, but that this flaw is minimized by Wheeler's
avoidance of the introduction of any quantitative geometry. As for
the mentioned quantum amplitudes, these need not have geometric
connotations as long as they do not arise from an inner product of state vectors. (See Table 2.)

\begin{table} \centering
\begin{tabular}{|l|c|c|} 
\hline
\textbf{Author} & \textbf{Basic objects} & \textbf{Basic relations}\\
\hline \hline
Clifford & feelings & nextness, succession \\ \hline
Eddington & \emph{points}, \emph{intervals} & statistical interaction \\ \hline
\multicolumn{3}{|l|}{Wheeler}\\ \hline
\quad Sewing machine & \emph{rings} & nextness (prob.~amp.) \\ \hline
\quad Borel set & \emph{points} & idem \\ \hline
\quad Calculus of propositions & propositions & statistical interaction \\ \hline
\quad Self-referential universe & acts of observation & idem \\ \hline
\end{tabular}
\caption{Summary of the use of basic objects and relations in the non-geometric or partly non-geometric approaches analyzed. Geometric concepts appear in italics. Abbreviations: Probability amplitude (prob.~amp.).}
\end{table}

Nagels' (1985) attempt started off with a scheme similar to
Wheeler's, where ``the only structure imposed on individual points
is a uniform probability of adjacency between two arbitrarily
chosen points'' (p.~545). He only assumed that these
probabilities are very small and that the total number of points
is very large (and possibly infinite). He thus believed his
proposal to satisfy the always desired requirement of ``a bare
minimum of assumptions'' and very natural ones at that. However,
much too soon he fell prey to quantitative geometry. Being a
pregeometric framework directly inspired in Wheeler, it is
startling to come across the following early remarks:
\begin{quote}
Without a background geometry, the simplest way to introduce
distance is to specify whether or not two given points are
``adjacent.'' [\ldots] We may then say that two ``adjacent''
points are a distance of 1 unit of length apart. (Nagels, 1985,
p.~546)
\end{quote}

For his part, Stuckey introduced in his attempt
\begin{quote}
a pregeometry that provides a metric and dimensionality over a
Borel set (Wheeler's ``bucket of dust'') without assuming
probability amplitudes for adjacency. Rather, a non-trivial metric
is produced over a Borel set $X$ per a uniformity base generated
via the discrete topological group structures over $X$. We show
that entourage multiplication in this uniformity base mirrors the
underlying group structure. One may exploit this fact to create an
entourage sequence of maximal length whence a fine metric
structure. (Stuckey, 2001, p.~1)
\end{quote}

Thus, both authors managed to create pregeometric metrics, i.e.\
metrics that will give a \emph{notion of distance for pregeometry}
and by means of which they hoped to obtain the usual spacetime
metric. This assertion is well supported by Stuckey and
Silberstein's (2000) remarks: ``our pregeometric notion of
distance'' (p.~9) and ``the process by which this metric yields a
spacetime metric with Lorentz signature must be obtained''
(p.~13). Stuckey (2001) moreover asserted that the pregeometries
of Nagels, Antonsen, and Requardt have, arguably, overcome the
difficulties represented by the presupposition of ``too much
geometric structure'' as previously brought up by Wheeler, since
their assumptions are minimal and ``the notion of length per graph
theory is virtually innate'' (p.~2). Arguably, indeed, because the
assumption of length is totally contrary to the spirit of
pregeometry according to its very inventor, and its introduction
acts to geometrically objectify what could have otherwise been
pregeometric schemes.

Alas, what has become of Wheeler's reasonable dictum: ``to admit
distance at all is to give up on the search for pregeometry,'' or
his demands for ``break[ing] loose at the start from all mention
of geometry and distance'' and for pregeometry to be ``as far
removed from geometry as the quantum mechanics of electrons is far
removed from elasticity?'' Or yet to quote Wheeler in two other
early, illuminating passages:
\begin{quote}
One might also wish to accept to begin with the idea of a
distance, or edge length, associated with a pair of these points,
even though this idea is already a very great leap, and one that
one can conceive of later supplying with a natural foundation of
its own.

[\ldots] [T]he use of the concept of distance between pairs of
points seems unreasonable\ldots [L]ength is anyway not a natural
idea with which to start. The subject of analysis here is
``pregeometry'', so the concept of length should be derived, not
assumed \emph{ab initio}. (Wheeler, 1964, pp.~497, 499)
\end{quote}
Has all this simply gone into oblivion?

The question, indeed, remains---why the need of a metric for
pregeometry? Should not pregeometry produce the traditional
spacetime metric by non-geometric means alone? Should it not go
even ``beyond Wheeler'' and do without any sort of geometric
objects too? This state of affairs comes to show most clearly
to what extent the craving of the human mind for geometric
explanations goes; at the same time, one catches a glimpse of the
difficulties that might be encountered in any attempt that
genuinely attempts to go beyond such explanations, so deeply
rooted in the mind.

After having spelt out Wheeler's vision for (his) pregeometry,
another reason supporting this venture is presented next, as we
near the end of this exploration. 

\subsection{The influence of language on thought}
Another context---supplementary to our observations of Section
\ref{PEST}---also exists for the belief that going beyond geometry
may be a necessary step for theoretical physics to take. Such a
context comes from the writings of linguist B.~L.~Whorf on his now
infamous principle of linguistic relativity. After a judicious
interpretation of it, one will be armed with quite a
straightforward idea about the influence of language on thought to
be applied in connection with the topic of this article. Whorf
wrote:
\begin{quote}
It was found that the background linguistic system\ldots of each
language is not merely a reproducing instrument for voicing ideas
but rather is itself the shaper of ideas\ldots We dissect nature
along lines laid down by our native languages. The categories and
types that we isolate from the world of phenomena we do not find
there because they stare every observer in the face; on the
contrary, the world is presented in a kaleidoscopic flux of
impressions which has to be organized by our minds---and this
means largely by the linguistic systems in our minds.

This fact is very significant for modern science, for it means
that no individual is free to describe nature with absolute
impartiality but is constrained to certain modes of interpretation
even while he thinks himself most free\ldots We are thus
introduced to a new principle of relativity, which holds that all
observers are not led by the same physical evidence to the same
picture of the universe, unless their linguistic backgrounds are
similar, or can in some way be calibrated. (Whorf, 1956,
pp.~212--214)
\end{quote}

One ought to be careful as to what to make of Whorf's remarks,
as their significance could go from (i) an implication of an
\emph{immutable constraint} of language upon thought with the
consequent fabrication of an unavoidable world picture to which
one is led by the language used, to (ii) milder insinuations about
language being ``only'' a \emph{shaper of ideas} rather than a
tyrannical master. How far is the influence of language on thought
to be taken to go?

In this, we will argue along with the ideas of Devitt and
Sterelny's. In their book, they clearly explained the
circular---although not viciously so---process in which thought
and language interact. On this issue, they wrote:
\begin{quote}
We feel a pressing need to understand our environment in order to
manipulate and control it. This drive led our early ancestors, in
time, to express a primitive thought or two. They grunted or
gestured, \emph{meaning something by} such actions. There was
speaker meaning without conventional meaning. Over time the grunts
and gestures caught on: linguistic conventions were born. As a
result of this trail blazing it is much easier for others to have
those primitive thoughts, for they can learn to have them from the
conventional ways of expressing them. Further, they have available
an easy way of representing the world, a way based on those
conventional gestures and grunts. They borrow their capacity to
think about things from those who created the conventions. With
primitive thought made easy, the drive to understand leads to more
complicated thoughts, hence more complicated speaker meanings,
hence more complicated conventions. (Devitt \& Sterelny, 1987,
p.~127)
\end{quote}

In the first place, this means that, as one could have already
guessed, thought must precede any form of language, or else how
could the latter have come into being? Secondly, it shows that
thought, as a source of linguistic conventions, is not in any way
restricted by language, although the characteristics of the latter
do, in fact, \emph{facilitate} certain forms of thought by making
them readily available, in the sense that the thinking processes
of many can now benefit from existing concepts already made by a
few others. This is especially the case in science, which abounds
in instances of this kind. For example, thought about complex
numbers is facilitated by the already invented imaginary-number
language convention ``$i=\sqrt{-1}$,'' just like the thought of a
particle being at different places at the same time is facilitated
by the already invented state-vector-related conventions of
quantum mechanics.

At the same time, an existing language can also \emph{discourage}
certain thoughts by making them abstruse and recondite to express
(Devitt \& Sterelny, 1987, p.~174). This does not mean that
certain things cannot be thought---they eventually can be---but
only that their expression does not come easily as it is not
straightforwardly supported by the language. As an example of this
instance, we can mention the reverse cases of the above examples.
That is to say, the evident difficulty of conjuring up any
thoughts about complex numbers and delocalized particles
\emph{before} the formal concepts above supporting these thoughts
were introduced into the language of science by some specific
individuals. The fact, however, that such linguistic conventions
were created eventually shows that thought beyond linguistic
conventions is possible---essential, moreover, for the evolution
of science.

One is thus led to the view that the picture that physics makes of
Nature is not only guided by the physicist's imagination, but also
by the physicist's language---with its particular vices and
virtues. This language, due to its very constitution attained
after a natural development, favours geometric modes of thinking,
while it appears to dissuade any sort of non-geometric
contemplation. Pregeometric schemes have shown this fact very
clearly: imagination was able to produce varied creative
frameworks, but they all spoke the same geometric language. Only
against this \emph{language-favoured background} did the
physicists' minds roam effortlessly.

The reason why humans have naturally developed primitive geometric
thoughts, leading to geometric modes of expression, in turn
reinforcing more geometric-like thinking, and so forth---in
conclusion, developing a geometric understanding---can only be
guessed at. Perhaps, it lies in the intricacies of their
evolutionary origins on Earth, and the resulting character and
dispositions of their brains.

The prospect of dispensing with geometric thinking is not
encouraging, for when geometry is lost, much is lost with it.
However, we feel this endeavour must be pursued. In this respect,
Misner et al.\ issued the following remarks:
\begin{quote}
And is not the source of any dismay the apparent loss of guidance
that one experiences in giving up geometrodynamics---and not only
geometrodynamics but geometry itself---as a crutch to lean on as
one hobbles forward? Yet there is so much chance that this view of
nature is right that one must take it seriously and explore its
consequences. Never more than today does one have the incentive to
explore [Wheeler's] pregeometry. (Misner et al., 1973, p.~1208)
\end{quote}

\section{Conclusion}
A long way has so far been travelled. Starting off in the realm of
geometry, its foundations in the form of geometric objects and
magnitudes were laid down and its use in all of present-day
physics was highlighted. It became evident that geometry and, in
particular, quantitative geometry is not only introduced into
theories to link them to experience via measurements, but also in
order to satisfy a more basic need seemingly possessed by the
human mind---a predilection to conceptualize the world after a
geometric fashion.

The abode of pregeometry, one of the latest attempts to throw new
light on the problem of space and time, was visited next. Although
judging by its name and goals pregeometry was the last place in
which one---for better or worse---would have expected to find
geometry, the former nevertheless fell prey to the latter.
Pregeometry, in its need for its own geometric means to explain
the origin of geometry in physics, cannot but raise the
disquieting question: is the human mind so dependent upon
geometric means of description that they cannot be avoided? Or has
pregeometry not tried hard enough? In any case, it must be
concluded that pregeometry has failed to live up to the semantic
connotations of its name and to the original intentions of its
creator, as well as to the intentions of its present-day
practitioners, only to become a considerable incongruity.

Next the vistas of a land farther beyond were surveyed. Different
older suggestions for the need to overcome geometry in physics
were analyzed, including those views pertaining to the father of
pregeometry proper. In general, this proposal stemmed from the
desire to reach into yet another layer in the nature of things.

On a whole, different related reasons were put forth for the need
to transcend geometry at the point where theoretical physics comes
to confront the unresolved problem of empty spacetime.
Essentially, the investigation of empty spacetime was seen to
demand ridding it of all its contents, including its metric field,
i.e.\ its quantitative geometry. Moreover, a look at the lessons
of history and a consideration of the mazelike nature of current
philosophical controversies suggested that understanding spacetime
anew demands going beyond its intrinsic geometric description
\emph{entirely}, for which the use of non-geometric methods is
required, as well as the guidance of a physical principle directly
relevant to the existence of spacetime. This conclusion was
supplemented by an investigation of the influence of language on
thought, which revealed that the language of physics may be
discouraging physicists' creative activities from effortlessly
straying into non-geometric realms.

The journey ends, for the time being, at a junction in the road at
which one must pause in order to evaluate what has been here
expounded---and make a decision. Geometry has furthered human
knowledge by a long way indeed, but has it come today to the
limits of its use? If the task of geometry has been completed, new
means should be devised that allow one to \emph{think} about
Nature---stimulated by a facilitation to \emph{talk} about
Nature---in a different, non-geometric way.

\section*{Acknowledgements} The authors are grateful to Antti
Tolvanen for valuable discussions.

\section*{References\footnote{Preprints of numerous published articles listed below are available at http://arxiv.org}}

\begin{enumerate}
\item Anandan, J.\ S.\ (1997). Classical and quantum physical
geometry. In R.\ S.\ Cohen, M.\ Horne, \& J.\ Stachel (Eds.),
\emph{Potentiality, entanglement and
passion-at-a-distance---Quantum mechanical studies for Abner
Shimony}, Vol.\ 2 (pp.\ 31--52). Dordrecht: Kluwer.

\item Antonsen, F.\ (1992). \emph{Pregeometry}. Master's Thesis,
University of Copenhagen, Niels Bohr Institute, 1992.

\item Ashtekar, A.\ (2005). Gravity and the quantum. \emph{New Journal
of Physics}, \emph{7}, 198 (1--33).

\item Aspect, A., Dalibard, J., \& Roger, G.\ (1982). Experimental
test of Bell's inequalities using time-varying analyzers. \emph{
Physical Review Letters}, \emph{49}, 1804--1807.

\item Bell, J.\ S.\ (1987). On the Einstein-Podolsky-Rosen
paradox. In \emph{Speakable and unspeakable in quantum mechanics}
(pp.\ 14--21). Cambridge: Cambridge University Press. (Original
work published 1964.)

\item Bombelli, L., Lee, J., Meyer, D., \& Sorkin, R.\ D.\ (1987).
Space-time as a causal set. \emph{Physical Review Letters},
\emph{59}, 521--524.

\item Boyer, C.\ B., \& Merzbach, U.\ C.\ (1991). \emph{A
history of mathematics}, second edition. New York: Wiley \& Sons.

\item Butterfield, J., \& Isham, C.\ (2000). Spacetime and the
philosophical challenge of quantum gravity. In C.\ Callender, \&
N.\ Huggett (Eds.), \emph{Physics meets philosophy at the Planck
scale} (pp.\ 33--89). Cambridge: Cambridge University Press.

\item Cahill, R.\ T., \& Klinger, C.\ M.\ (1996).
Pregeometric modelling of the spacetime phenomenology.
\emph{Physics Letters A}, \emph{223}, 313--319.

\item Cahill, R.\ T., \& Klinger, C.\ M.\ (1997).
Bootstrap universe from self-referential noise.
arXiv:gr-qc/9708013 v1.

\item Clifford, W.\ K.\ (1886).
The Unseen Universe. In L.\ Stephen, \& F.\ Pollock (Eds.),
\emph{Lectures and essays} (pp.\ 161--179). London: Macmillan \&
Co. (Original work published 1875.)

\item Dadi\'{c}, I.\ and Pisk, K.\ (1979).
Dynamics of discrete-space structure. \emph{International Journal
of Theoretical Physics}, \emph{18}, 345--358.

\item Demaret, J., Heller, M., \& Lambert, D. (1997). Local and global properties of the
world. \emph{Foundations of Science}, \emph{2}, 137--176.

\item Devitt, M., \& Sterelny, K.\ (1987). \emph{Language and
reality: An introduction to the philosophy of language}. Oxford:
Basil Blackwell.

\item Dirac, P.\ A.\ M.\ (1958). \emph{The principles of quantum mechanics}, fourth edition. Oxford: Clarendon Press.

\item Eakins, J., \& Jaroszkiewicz, G.\ (2002).
The quantum universe. arXiv:quant-ph/0203020 v1.

\item Eakins, J., \& Jaroszkiewicz, G.\ (2003). The origin of causal
set structure in the quantum universe. arXiv:gr-qc/0301117 v1.

\item Eddington, A.\ (1920). \emph{Space, time and
gravitation---An outline of the general relativity theory}.
Cambridge: Cambridge University Press.

\item Einstein, A.\ (1952). \emph{Relativity---The special and the general
theory}, 15th edition. New York: Three Rivers Press.

\item Gibbs, P.\ (1996). The small scale structure of space-time:
A bibliographical review. arXiv:hep-th/9506171 v2.

\item Hill, E.\ L.\ (1955). Relativistic theory of
discrete momentum space and discrete space-time. \emph{Physical
Review}, \emph{100}, 1780--1783.

\item Horzela, A., Kapu\'{s}cik, E.,
Kempczy\'{n}ski, J., \& Uzes, C.\ (1992). On discrete models of
space-time. \emph{Progress in Theoretical Physics}, \emph{88},
1065--1071.

\item Isham, C.\ (1995). \emph{Quantum theory: Mathematical and structural
foundations}. London: Imperial College Press.

\item Jaroszkiewicz, G.\ (2001). The running of the universe and
the quantum structure of time. arXiv:quant-ph/0105013 v2.

\item Lehto, M.\ (1988). \emph{Simplicial quantum gravity}.
Doctoral thesis, Department of Physics, University of
Jyv\"{a}skyl\"{a}, Research report No.\ 1/1988.

\item Lehto M., Nielsen H.\ B., \& Ninomiya, M.\ (1986a).
Pregeometric quantum lattice: A general discussion. \emph{Nuclear
Physics B}, \emph{272}, 213--227.

\item Lehto M., Nielsen H.\ B., \& Ninomiya, M.\ (1986b).
Diffeomorphism symmetry in simplicial quantum gravity.
\emph{Nuclear Physics B}, \emph{272}, 228--252.

\item Misner, C.\ W., Thorne, K.\ S., \& Wheeler, J.\ A.\ (1973).
\emph{Gravitation}. New York: Freeman \& Co.

\item Monk, N.\ A.\ M.\ (1997). Conceptions of space-time:
Problems and possible solutions. \emph{Studies in History and
Philosophy of Modern Physics}, \emph{28}, 1--34.

\item Nagels, G.\ (1985). Space as a ``bucket of dust.'' \emph{General
Relativity and Gravitation}, \emph{17}, 545--557.

\item Ohanian, H.\ C.\ (1976). \emph{Gravitation and spacetime}. New York: Norton \& Co.

\item Patton, C.\ M., \& Wheeler, J.\ A.\ (1975). Is physics legislated by
cosmogony? In C.\ J.\ Isham, R.\ Penrose, \& D.\ W.\ Sciama
(Eds.), \emph{Quantum gravity, an Oxford symposium} (pp.\
538--605). Oxford: Clarendon Press.

\item Perez Bergliaffa S.\ E., Romero, G.\ E., \& Vucetich,
H.\ (1998). Towards an axiomatic pregeometry of space-time.
\emph{International Journal of Theoretical Physics}, \emph{37},
2281--2298.

\item Requardt, M.\ (1995). Discrete mathematics and
physics on the Planck-scale. arXiv:hep-th/9504118 v1.

\item Requardt, M.\ (1996). Emergence of space-time on the Planck scale
described as an unfolding phase transition within the scheme of
dynamical cellular networks and random graphs.
arXiv:hep-th/9610055 v1.

\item Requardt, M.\ (2000).  Let's call it nonlocal quantum
physics. arXiv:gr-qc/0006063 v1.

\item Requardt, M., \& Roy, S.\ (2001). (Quantum) space-time
as a statistical geometry of fuzzy lumps and the connection with
random metric spaces. \emph{Classical and Quantum Gravity},
\emph{18}, 3039-3058.

\item Riemann, B.\ (1873). \"{U}ber die Hypothesen, welche
der Geometrie zu grunde liegen. In W. K.\ Clifford (Trans.),
\emph{Nature}, \emph{VIII} (183, 184), 14--17, 36, 37. (Original
work published 1867.)

\item Selleri, F.\ (1988). \emph{Quantum mechanics versus local realism---The
Einstein-Podolski-Rosen paradox}, F.\ Selleri (Ed.). New York:
Plenum Press.

\item Sorkin, R.\ D.\ (1991). Spacetime and causal
sets. In J.\ C.\ D'Olivo, E.\ Nahmad-Achar, M.\ Rosenbaum, M.\ P.\
Ryan, L.\ F.\ Urrutia, \& F.\ Zertuche (Eds.), \emph{Relativity
and gravitation: Classical and quantum} (pp.\ 150--173).
Singapore: World Scientific.

\item Sorkin, R.\ D.\ (2005). Causal sets: Discrete
gravity. In  A.\ Gomberoff, \& D.\ Marolf (Eds.),
\emph{Lectures on quantum gravity} (Series of the Centro de
Estudios Cient\'{i}ficos). Berlin: Springer.

\item Stapp. H.\ P. (1975). Bell's theorem and world process. \emph{Il Nuovo
Cimento B}, \emph{29}, 270--276.

\item Stuckey, W.\ M.\ (2001). Metric structure and dimensionality over a
Borel set via uniform spaces. arXiv:gr-qc/0109030 v2.

\item Stuckey, W.\ M., \& Silberstein, M.\ (2000). Uniform spaces in the pregeometric modelling
of quantum non-separability. arXiv:gr-qc/0003104 v2.

\item Van der Waerden, B.\ L.\ (1983). \emph{Geometry and algebra in ancient
civilizations}. Berlin: Springer.

\item Volovich, I.\ V.\ (1987). Number theory as the ultimate
physical theory. Preprint CERN-TH., 4781--4787.

\item Wheeler, J.\ A.\ (1964). Geometrodynamics and the issue of the
final state. In C.\ De Witt, \& B.\ S.\ De Witt (Eds.),
\emph{Relativity, groups and topology} (pp.\ 317--520). New York:
Gordon \& Breach.

\item Wheeler, J.\ A.\ (1980). Pregeometry: Motivations and prospects.
In \ A.\ R.\ Marlov (Ed.), \emph{Quantum theory and gravitation}.
New York: Academic Press.

\item Whorf, B.\ L.\ (1956). Science and linguistics. In
J. B.\ Carroll (Ed.), \emph{Language, thought and reality} (pp.\
207--219). Cambridge, MA: M.I.T Press.

\item Wilson, R.\ J.\ (1985). \emph{Introduction to graph
theory}, third edition. Essex: Longman Group.
\end{enumerate}

\end{document}